# Multi-UTE Myelin Fraction Mapping: Repeatability and Cross-Field-Strength Reproducibility at 1.5T and 3T

Nan Yin[1], Marco Reisert[1], Theo Demerath[2], Shuai Liu[1], Alexander Rau[2], Uzay Emir[3,4], Horst Urbach[2], Michael Bock[1], Ali Caglar Özen[1]

[1]Division of Medical Physics, Department of Diagnostic and Interventional Radiology, University Medical Center Freiburg, Faculty of Medicine, University of Freiburg, Freiburg, Germany

[2]Department of Neuroradiology, Faculty of Medicine, University of Freiburg, Freiburg, Germany

[3]Department of Radiology, University of North Carolina, Chapel Hill, NC, USA

[4]The Lampe Joint Department of Biomedical Engineering, University of North Carolina, Chapel Hill, NC, USA

Corresponding author: Nan Yin, M.Sc.

University of Freiburg, Freiburg, Germany

Killianstraße 5a, 79106 Freiburg

Telephone: +49 761 270-71780

nan.yin@uniklinik-freiburg.de

Word count (main text): 5000/5000

# Abstract

**Purpose:** To evaluate the repeatability and cross-field-strength reproducibility (1.5T vs 3T) of *in vivo* cerebral myelin fraction (MF) mapping with multi-contrast dual-resolution 3D UTE (multi-UTE), and to compare it with magnetization transfer (MT)-based methods.

**Methods:** MT methods provide an indirect myelin-sensitive measure derived from magnetization exchange between macromolecular and free-water proton pools, whereas multi-UTE targets the ultrashort- $T_2^*$ signal components. In five healthy volunteers, multi-UTE and MT-based measurements were performed at 1.5T and 3T, and the calculated MF maps were compared with magnetization transfer ratio (MTR) and magnetization transfer saturation (MTsat) maps. Two volunteers underwent repeated multi-UTE measurements within and between days. Field-strength reproducibility and test-retest repeatability were evaluated using linear regression, Bland-Altman analysis, concordance correlation coefficient (CCC), and intraclass correlation coefficient (ICC).

**Results:** Multi-UTE enabled simultaneous acquisition of quantitative and high-resolution structural images at both 1.5T and 3T. MF repeatability was high at 3T (ICC = 0.94-0.96) and more variable at 1.5T (0.64-0.96). Across field strengths, MF and MTsat showed comparable reproducibility (CCC 0.70 vs 0.60) with small median absolute differences (8.8% vs 8.9%). MTsat showed tighter correlation ($R^2$ 0.77 vs 0.68), while MF showed a wider inter-regional dynamic range. MTR differed strongly between field strengths (CCC = 0.06).

**Conclusion:** Multi-UTE MF, obtained without $B_1$ or $T_1$ mapping and alongside a high-resolution structural image, achieves cross-field reproducibility comparable to, but not exceeding, MTsat. Lower SNR at 1.5T and unmodelled $T_1$ weighting may have contributed to residual field dependence.

# 1. INTRODUCTION

Myelin is a lipid-rich, multilayered membrane that encases axons in the central nervous system. Myelin facilitates fast and effective neuronal transmission via saltatory conduction; myelinating oligodendrocytes also play a role in maintaining long-term axonal integrity, providing metabolic support, and ensuring tissue homeostasis[1,2]. Myelin formation and remodeling are integral to normal neurodevelopment, whereas demyelination is associated with a wide range of neurological diseases, including multiple sclerosis[1,2]. Accurate in vivo assessment of cerebral myelin is crucial for characterizing typical white-matter architecture, identifying pathological changes, and monitoring long-term disease progression or treatment responses[3].

Several quantitative MRI techniques have been developed to characterize myelin-related tissue properties in vivo. Conventional myelin imaging methods include magnetization-transfer (MT)-based metrics, myelin water imaging (MWI), and susceptibility-based techniques such as χ-separation. MT-based methods in particular probe the magnetization exchange between mobile water protons and the semisolid macromolecular proton pool which is altered in patients with demyelinating diseases[4]. Compared with the conventional magnetization transfer ratio (MTR), magnetization transfer saturation (MTsat) reduces sensitivity to $T_1$ relaxation and radiofrequency-field inhomogeneity, and has been tested in multi-center clinical studies[5–9]. MWI instead probes the rapidly relaxing water pool confined between myelin bilayers and reports its contribution as the myelin water fraction (MWF), which has shown quantitative correspondence with histological myelin staining[10,11]. Recent systematic reviews have reported substantial variability in both histological correspondence and measurement reproducibility across techniques and studies[3,12].

Ultrashort echo time (UTE) and zero echo time (ZTE) sequences offer a more direct myelin quantification by targeting the ultrashort-$T_2^*$ proton pools. Despite the blurring associated with ultrashort-$T_2^*$ values on a sub-millisecond timescale, studies have shown that the residual signal detected with UTE and ZTE arises predominantly from non-aqueous myelin protons and tightly bound water at the myelin bilayer[13–16]. Variants of UTE and ZTE sequences have also been evaluated in myelin phantoms, ex vivo, and in vivo studies[15,17–22].

Recently, we introduced a multi-contrast dual-resolution 3D UTE sequence (multi-UTE) that combines high-resolution water-excited structural imaging with interleaved multi-echo UTE acquisitions in a single scan[23]. This sequence measures the ultrashort-$T_2^*$ component myelin fraction (MF), which correlated with electron-microscopy-derived myelin volume fraction ($R^2$ = 0.86) in ex vivo sheep brain and enabled in vivo cerebral MF mapping at 3T. However, this initial study was performed at a single field strength and did not characterize the repeatability of the MF quantification. If MF reflects a tissue fraction, it should be largely field-independent. We assessed the repeatability of the apparent MF measured with multi-UTE and compared its reproducibility between 1.5T and 3T with that of MTsat.

This study evaluated the technical feasibility and precision of multi-UTE-derived MF in healthy volunteers at 1.5T and 3T as most widely available clinical field strengths. We assessed within-scanner test-retest repeatability separately at each field strength and cross-field-strength reproducibility, following established quantitative-imaging metrology terminology[24]. MTsat served as the benchmark comparison metric as it is based on a standard 3D FLASH sequence, which is readily available on clinical MRI systems and was validated in a large multi-center patient cohort[8,9]. MTR was used as an uncorrected baseline. For concurrent comparison, regional correspondence between MF using multi-UTE, MTR and MTsat was quantified at each field strength.

# 2. METHODS

## 2.1 MRI acquisition and calculation

The multi-UTE sequence[23] uses a 3D radial k-space trajectory (Fig. 1). Two low-resolution echoes, $S_1$ and $S_2$, were acquired between the two pulses of a 1-1 binomial water-excitation pulse pair using balanced readout gradients. The inter-pulse interval was defined as $TR_1 = 1/(2 \cdot \Delta f)$, where $\Delta f$ is the fat-water frequency difference. Accordingly, $TR_1$ was increased from 1.19 ms at 3T to 2.38 ms at 1.5T to account for the lower fat-water frequency separation and preserve fat suppression. After the second excitation pulse, a high-resolution UTE image, $S_3$, was acquired at an independently defined $TE_3$, followed by a fixed spoiler gradient for each spoke. $TE_1$ and $TE_2$ were varied across six acquisitions ($N_{acq}$ = 6) to characterize the ultrashort-$T_2^*$ component attributed to myelin. At 1.5T, two TE distributions were used (TE distribution 1 used the same TE values as the 3T protocol, whereas TE distribution 2 covered a wider TE range. Details are given in Table 1). $S_3$ contained six times as many spokes as the $S_1$ or $S_2$ dataset ($N_{spokes,S3}$=$N_{acq}$×$N_{spokes,S1}$=6×40,000=240,000).

Numerical signal calculations were performed to characterize the field- and off-resonance-dependent responses of $S_1$–$S_3$ according to the multi-UTE signal equation[23] using literature values[15,16,25–28]:

$$S_1 = M_z^{SS} \cdot \sin\alpha \cdot e^{-TE_1/T_2^*} \quad [1]$$

$$S_2 = M_z^{SS} \cdot \sin\alpha \cdot e^{-TE_2/T_2^*} \quad [2]$$

$$S_3 = \sqrt{\left\{M_z^0\left(1 - e^{-\frac{TR_1}{T_1}}\right) + M_z^{SS}\cos\alpha\left(e^{-\frac{TR_1}{T_1}} + \cos(\Delta\Psi)\, e^{-\frac{TR_1}{T_2^*}}\right)\right\}^2 + \left\{M_z^{SS}\sin(\Delta\Psi)\, e^{-\frac{TR_1}{T_2^*}}\right\}^2} \cdot \sin\alpha \cdot e^{-TE_3/T_2^*} \quad [3]$$

, where

$$M_z^{SS} = M_z^0 \cdot \frac{1 - e^{-\frac{TR_2}{T_1}} + \left(1 - e^{-\frac{TR_1}{T_1}}\right)\cdot\left(\cos\alpha \cdot e^{-\frac{TR_2}{T_1}}\right)}{1 - \cos^2\alpha \cdot e^{-\frac{TR_1+TR_2}{T_1}} + \sin^2\alpha \cdot \cos\Delta\Psi \cdot e^{-\frac{TR_2}{T_1}} \cdot e^{-\frac{TR_1}{T_2^*}}} \quad [4]$$

,$\Delta\Psi$ is the off-resonance phase at $TR_1$ before the second $\alpha$ pulse ($\Delta\Psi = TR_1 \cdot \Delta\omega$).

Multi-UTE images were reconstructed using re-gridding with a Kaiser-Bessel convolution kernel, followed by a 3D Hamming filter, density compensation, 3D fast Fourier transformation, deapodization, followed by ESPIRiT[29] coil combination, implemented in MATLAB (MathWorks, MA, USA).

The multi-echo signal model is:

$$S(TE) = \left[S_{0,us} \cdot \mathrm{e}^{\mathrm{i}(\Delta\omega_{us}TE)} \cdot \mathrm{e}^{-\frac{TE}{T_{2,us}^*}} + S_{0,s} \cdot \mathrm{e}^{\mathrm{i}(\Delta\omega_{s}TE)} \cdot \mathrm{e}^{-\frac{TE}{T_{2,s}^*}} + S_{0,l}\right] \cdot \mathrm{e}^{\mathrm{i}\phi} \qquad [5]$$

, where $\phi$ is an arbitrary initial phase, $\Delta\omega_{us}$ and $\Delta\omega_s$ are off-resonance frequency shifts of ultrashort-$T_{2,us}^*$ and short-$T_{2,s}^*$ components, respectively. The long-$T_2^*$ (free-water) component was assumed constant across the ultrashort-TE range studied, given less than 0.5% decay over this timescale. MF was defined as $MF = S_{0,us}/(S_{0,us} + S_{0,s} + S_{0,l})$. Model fitting was performed using a principal-component-analysis (PCA)-based Bayesian estimator[23,30]. The full parameter range and uncertainty simulation of the Bayesian estimator are given in Supporting Information Figure S1.

## 2.2 Multi-parameter-mapping (MPM) acquisition

The MPM protocol comprised three multi-echo 3D FLASH acquisitions with predominant $T_1$, proton-density, and magnetization-transfer weighting, hereafter referred to as T1w, PDw and MTw, respectively. At each field strength, the PDw and MTw acquisitions were matched for spatial resolution, repetition time, flip angle, and echo times. Table 1 summarizes the field-strength-specific acquisition parameters. For the MTw acquisition, magnetization-transfer contrast was generated using a sinc-shaped MT pulse with a frequency offset of 1200 Hz and a duration of 10 ms. No MT preparation pulse was applied during the corresponding PDw acquisition.

Image data were then processed using the hMRI toolbox[7–9,31] in SPM[32]. With full ordinary-least-squares processing and UNICORT $B_1$ correction enabled, the multi-echo PDw, T1w, and MTw signals were fitted using ESTATICS, and the contrast-specific signal intercepts extrapolated to TE=0 were stored in a map. Because the PDw and MTw acquisitions had identical TRs and flip angles, the magnetization transfer ratio (MTR) was calculated voxel-wise as:

$$MTR = (S_{PDw} - S_{MTw})/S_{PDw} \ . \qquad [6]$$

Magnetization transfer saturation was calculated according to the Helms model implemented in the hMRI toolbox:

$$MTsat = (\mathrm{A}\alpha_{MT}/S_{MTw} - 1)R_1TR_{MT} - \alpha_{MT}^2/2 \ , \qquad [7]$$

where A is the unnormalized signal amplitude of the PDw and T1w data, $R_1$ is the longitudinal relaxation rate, and $\alpha_{MT}$ and $TR_{MT}$ are the MTw excitation flip angle and repetition time, respectively.

## 2.3 MRI measurements

The sequences were tested on clinical 3T and 1.5T MRI systems (MAGNETOM Prisma and Sola, Siemens, Erlangen, Germany). A 20-channel head coil array (Head/Neck 20, Siemens) was used for signal reception. To minimize head motion, an inflatable head support (Pearl Technology, Switzerland) was placed between the head and the coil. The study was conducted in accordance with the Declaration of Helsinki and approved by the Ethics Committee of the University of Freiburg (No. 160/2000). Five healthy volunteers (4 males, 1 female; age range 26–38 years) were included. All participants provided written informed consent prior to the measurements.

All volunteers were measured using the same multi-UTE protocol at both 3T and 1.5T to enable within-subject comparison of the MF between field strengths. Two of five volunteers underwent additional multi-UTE acquisitions for test-retest evaluation under three conditions: same day repetition, different-day repetition, and repetition with altered acquisition parameters (two TE distributions at 1.5T). MPM data were additionally acquired at both field strengths to generate MTR and MTsat for comparison with the multi-UTE-derived MF. T1-weighted magnetization-prepared rapid gradient-echo (T1-MPRAGE) images were acquired as an anatomical reference. An overview of the acquisition protocols and sequence parameters is provided in Table 1.

## 2.4 Data analysis and evaluation

To visualize signal decay over the ultrashort-TE range, a voxel-wise signed difference image was calculated:

$$\Delta S_1 = S_1(TE_1 = 70\ \mu s) - S_1(TE_1 = 250\ \mu s) \qquad [8]$$

Voxel-wise $B_0$ maps were calculated from the complex multi-UTE images acquired at two echo times:

$$\Delta B_0 = \frac{\arg\left[I(TE_2)\cdot I^*(TE_1)\right]}{-2\pi(TE_2 - TE_1)} \qquad [9]$$

Here, $I$ is the complex MR signal, and * denotes complex conjugation.

Image processing was performed using the software packages FSL[33] and FreeSurfer[34]. The 3T high-resolution multi-UTE $S_3$ image was segmented with SynthSeg[35] and used as anatomical reference. T1-MPRAGE images were also acquired to benchmark multi-UTE $S_3$. Registration with SynthMorph[36] mapped the FreeSurfer and JHU ICBM-DTI atlas labels to each participant's anatomy. Quantitative maps remained in their native grids, while inverse transformations propagated the 20 largest JHU ROIs and 14 FreeSurfer ROIs to these grids using nearest-neighbor interpolation. ROIs generated using both reference strategies were overlaid on the lower-resolution $S_1$ of multi-UTE image for quality control. Visual quality control and quantitative comparisons of the two strategies are shown in the Supporting Information Figure S2.

All statistical analyses were performed in MATLAB (MathWorks, MA, USA). For each atlas (FreeSurfer and JHU DTI), regional MF, MTR, and MTsat were summarized as mean values within each ROI per subject and acquisition. Agreement between paired measurements was assessed using linear regression (slope, intercept, $R^2$, and Spearman's rank correlation $\rho$), Bland-Altman analysis (mean bias and 95% limits of agreement), concordance correlation coefficient (CCC), and intraclass correlation coefficient (ICC, two-way random-effects, absolute agreement, single measures). Differences were assessed using two-sided paired t-tests. Statistical significance was set at $p<0.05$.

Multi-UTE repeatability was evaluated in two volunteers using ROI-wise MF values obtained under two repeated-acquisition conditions: same-day repetition, and different-day repetition. Cross-field-strength reproducibility was assessed by comparing paired ROI-wise MF values obtained at 3T (TR=7ms protocol) with those obtained at 1.5T using each of the two TE distributions. To directly benchmark field-strength robustness, MF (from multi-UTE), MTR and MTsat (from 3D FLASH MT) were each compared between 1.5T and 3T using linear regression, Bland-Altman bias and limits of agreement, root-mean-square error (RMSE), normalized RMSE, median absolute symmetric percentage difference, CCC, and ICC, computed separately. The regional association between MF and MTsat was evaluated separately at 1.5T and 3T using linear regression and Bland-Altman analysis. For regional analysis, the mean MF was calculated within each selected region of interest (ROI) defined by the FreeSurfer and JHU DTI atlases, and the resulting values were compared between 1.5T and 3T.

# 3. RESULTS

As shown in Figure 2, the small chemical-shift offsets of WM and GM resulted in minor changes in the multi-UTE signal curves at 1.5T and 3T only. As expected, $S_3$ yielded good WM-GM contrast with signal amplitudes about twice as large as $S_1$. Although the dominant myelin resonance was near the lipid resonance, its ultrashort-$T_2^*$ caused the transverse magnetization to decay almost completely by longer TEs ($S_2$). Therefore, the binomial fat-suppression pulse had little effect on the myelin signal, and the $S_3$ of the myelin pool was close to a single-RF-pulse excitation. In addition, at a low FA of 4°, the field-dependent $T_1$ and $T_2^*$ values result in a simulated MF that is only 1% lower (6.7% relative) at 1.5T than at 3T (Supporting Information Figure S7). Agreement analysis of the Bayesian estimates (Supporting Information Figure S1) showed that, at matched SNRs, MF estimates at 1.5T tracked the identity line more closely than those at 3T over the simulated range. Normalizing $S_{0,us}$ by the total signal produced a modest broadening of the prediction distribution and slightly reduced accuracy, reflecting limited TE coverage and poorer identifiability of long $T_2^*$ components.

Figure 3 compares in vivo multi-UTE signals of ROIs (mean$\pm$std) at 3T with 1.5T signals acquired using two TE distributions. As expected, increasing TE resulted in decreased signal intensity in every ROI and was consistently higher at 3T than at 1.5T. WM showed higher signal intensity than GM, consistent with the T1-weighted contrast and simulation results. The WM curves generally showed larger absolute signal decreases than the GM curves. MF was then estimated from the mean signal across all included voxels in each ROI. The L-CWM MF estimate was significantly higher than those of the selected GM regions at both field strengths (P < 0.01). Across the JHU tracts, TE distribution 1 yielded slightly lower MF values at 1.5T than at 3T (0.081–0.092 versus 0.092–0.099). In contrast, TE distribution 2 at 1.5T yielded slightly higher MF (0.096–0.108).

Figure 4 summarizes the integrated multi-UTE results at 3T and 1.5T. The method simultaneously yielded ultrashort-$T_2^*$ signal fraction maps and high-resolution anatomical images with high contrast-to-noise ratio near 30 between WM and GM. The comparable outputs obtained at both field strengths demonstrate the cross-field feasibility and consistency of our proposed approach.

The 3T results are shown in Figure 4a. From left to right, the panels present the low-resolution magnitude images $S_1(TE_1 = 70\ \mu s)$ and $S_1(TE_1 = 250\ \mu s)$, the subtraction image $\Delta S_1$, the ultrashort-$T_2^*$ signal fraction map, the Bayesian-estimation error map, and the high-resolution multi-UTE image. The lipid-rich scalp showed high signal intensity in $S_1(TE_1 = 70\ \mu s)$. The subtraction image highlighted ultrashort-$T_2^*$ signals from cranial bone, scalp fat, and myelin-associated components within the brain. In contrast, air and the water signal in muscle and cerebrospinal fluid appeared hypointense, consistent with effective removal of long-$T_2^*$ components. The subtraction image appeared more blurred because of PSF broadening. Within the brain, the ultrashort-$T_2^*$ signal fraction map provided myelin-sensitive contrast. MF values exceeded 0.08 in WM and reached ≥0.1 in major central fiber tracts, and MF was smaller than 0.07 in GM regions such as the putamen and thalamus. The Bayesian-estimation error was low throughout most of the brain, indicating close agreement between the measured and model-predicted signals. Inaccuracies increased in the air and near the head-support apparatus, where the estimations showed greater uncertainty. The high-resolution $S_3$ image provided strong T1-weighted anatomical contrast. Fat suppression reduced the bright scalp-fat signal and further improved tissue contrast. The multi-UTE $S_3$ images yielded excellent anatomical contrast and may serve as an alternative to clinical T1-weighted MPRAGE for structural imaging and tissue segmentation, with segmentation results closely matching those derived from T1w-MPRAGE; a detailed comparison is provided in the Supporting Information Figure S2.

Figure 4b shows the corresponding multi-UTE results at 1.5T. Ultrashort-$T_2^*$ components remained clearly visible in both the subtraction image and the signal-fraction map. The intracranial distribution was consistent with that observed at 3T, with higher ultrashort-$T_2^*$ fractions in WM than in GM. In white matter, the SNR of multi-UTE $S_1$ was 568 at 3T and 337 at 1.5T, representing a 40% reduction at 1.5T and consequently greater Bayesian-estimation uncertainty. As expected at the lower field strength, the high-resolution $S_3$ multi-UTE images acquired at 1.5T also appeared slightly noisier than those acquired at 3T.

Test–retest analysis showed strong same-day repeatability of MF measurements (Fig. 5a, b). Repeated measurements closely followed the identity line ($y = 0.959x + 0.002$, $R^2 = 0.994$; Spearman's $\rho = 0.989$). Bland–Altman analysis showed a small negative mean difference (bias) of −0.0021 and narrow 95% limits of agreement (LoA, −0.0042 to 0.0000). These findings indicate good short-term repeatability with a small negative within-day shift. Across days, MF measurements remained strongly correlated ($y = 1.032x - 0.006$, $R^2 = 0.986$; Spearman's $\rho = 0.992$; Fig. 5c). Bland–Altman analysis revealed a slightly larger negative mean difference of −0.0031, with 95% LoA from −0.0061 to 0.0000 (Fig. 5d). Thus, the relative ordering of regional MF values was well preserved across days, whereas their absolute values showed a systematic inter-day offset.

Cross-field-strength agreement was evaluated in JHU DTI atlas ROIs by comparing values measured at 3T and 1.5T (Fig. 6). For multi-UTE MF, values measured at 3T with TR = 7 ms were compared with those measured at 1.5T using two echo-time configurations. The TE distribution 1 produced a regression slope close to unity ($y = 1.013x + 0.007$, $R^2 = 0.745$; Spearman's $\rho = 0.791$). Bland–Altman analysis showed a positive bias of 0.0082 (95% LoA, −0.0046 to 0.0210), indicating that MF values measured at 3T were, on average, higher than those measured at 1.5T (Fig. 6a, b). The TE distribution 2 showed a higher correlation ($y = 1.136x - 0.019$, $R^2 = 0.807$; Spearman's $\rho = 0.871$), although its fitted slope deviated further from unity. The mean bias was −0.0067 (95% LoA, −0.0181 to 0.0047), indicating that MF values measured at 3T were, on average, lower than those measured at 1.5T (Fig. 6c, d). Comparisons of MTR and MTsat values measured at both field strengths showed contrasting patterns. MTR demonstrated a weaker association ($y = 1.388x + 0.027$, $R^2 = 0.481$; Spearman's $\rho = 0.589$). The positive bias of 0.1646 (95% LoA, 0.1488–0.1804) indicated that MTR values measured at 3T were systematically higher than those measured at 1.5T (Fig. 6e, f). In contrast, MTsat showed a stronger association and a regression line close to identity ($y = 1.033x - 0.004$, $R^2 = 0.789$; Spearman's $\rho = 0.878$). Its negative bias of −0.0023 (95% LoA, −0.0057 to 0.0011) indicated that MTsat values measured at 3T were, on average, slightly lower than those measured at 1.5T (Fig. 6g, h).

Table 2 summarizes the test–retest repeatability and cross-field-strength reproducibility of multi-UTE MF at 3T and 1.5T. At 3T, MF showed consistently high repeatability in both participants. Slopes ranged from 0.93 to 1.00, $R^2$ from 0.97 to 0.98, CCC from 0.94 to 0.96, and ICC from 0.94 to 0.96. RMSE values ranged from 0.0031 to 0.0040, and median absolute symmetric differences ranged from 2.8% to 4.3%. At 1.5T, repeatability varied across TE protocols and participants. Across both protocols, slopes ranged from 0.95 to 1.10, $R^2$ from 0.89 to 0.95, CCC from 0.63 to 0.95, and ICC from 0.64 to 0.96. RMSE values ranged from 0.0029 to 0.0088, and median absolute symmetric differences ranged from 2.9% to 10%.

MF showed better cross-field-strength reproducibility than MTR across all reported metrics with slopes of 0.98 and 0.80, $R^2$ values of 0.68 and 0.23, and RMSE of 0.0090 and 0.089, respectively. MF also showed lower RMSE/ROI SD (0.86 versus 14) and median absolute symmetric difference (8.8% versus 21%), together with higher CCC (0.70 versus 0.06) and ICC (0.71 versus 0.06). Compared with MF, MTsat had higher $R^2$ (0.77 versus 0.68) and lower RMSE (0.0040 versus 0.0090), but lower CCC (0.60 versus 0.70) and ICC (0.60 versus 0.71). Overall, MF was more reproducible than MTR, whereas its performance relative to MTsat depended on the metric.

Regional MF derived using the same multi-UTE method showed broadly consistent distributions between 1.5T and 3T under both atlas-based ROI definitions (Fig. 7). In the FreeSurfer analysis, the regional ordering was preserved across field strengths. Cerebral white matter showed the highest MF, the amygdala and hippocampus showed lower values, and the remaining subcortical regions showed intermediate values. The violin distributions and interquartile ranges overlapped substantially between 1.5T and 3T for most ROIs. Across field strengths, cerebral white matter had higher MF than the combined subcortical ROIs ($p < 0.01$), while the pallidum–putamen–thalamus group had higher MF than the amygdala–caudate–hippocampus group ($p < 0.05$). The JHU atlas showed a similarly preserved regional pattern at both field strengths. The GCC and PLIC generally exhibited lower MF, whereas the ACR, PCR, RLIC, SCR, and SLF showed higher values. Despite modest field-dependent shifts in several tracts, the distributions and interquartile ranges at 1.5T and 3T remained substantially overlapping. Overall, the two atlas-based analyses showed high cross-field consistency of regional multi-UTE MF measurements.

# 4. DISCUSSION

In this study, we adapted the multi-UTE sequence to 1.5T and characterized the repeatability and cross-field-strength reproducibility of the multi-UTE-derived myelin fraction (MF) in healthy volunteers at 1.5T and 3T. MF showed high test-retest agreement at 3T (ICC = 0.94-0.96) and more variable agreement at 1.5T (ICC = 0.64-0.96). Across-field strengths, MF showed moderate agreement (CCC = 0.70, ICC = 0.71), with a median absolute symmetric difference of 8.8%. This was comparable to MTsat, which corrects explicitly for $T_1$ relaxation and $B_1$ inhomogeneity (CCC = 0.60, median absolute symmetric difference 8.9%). MTsat showed a tighter linear association between field strengths ($R^2$ = 0.77 vs 0.68), whereas MF showed a wider inter-regional dynamic range. Uncorrected MTR, by contrast, differed substantially between field strengths (CCC = 0.06). Multi-UTE thus provides a myelin-sensitive quantitative metric with field-strength robustness similar to that of an established, correction-based MT metric, without requiring separate $B_1$ or $T_1$ mapping, while simultaneously delivering a high-resolution structural image and $B_0$ information.

To our knowledge, this is the first in vivo quantification of the ultrashort-$T_2^*$ signal fraction in the human brain at 1.5T. Adapting the sequence to the lower field strength required doubling the inter-pulse interval of the binomial water-excitation pulse pair ($TR_1$ = 2.38 ms) to match the smaller fat–water frequency separation. This longer interval benefits the multi-UTE concept: it extends the time available for the interleaved low-resolution echoes and allows a wider TE coverage within each acquisition, as used in TE distribution 2. The main cost is the lower SNR at 1.5T, which reduced the $S_1$ SNR in white matter by about 40% and increased the uncertainty of the Bayesian estimates (Supporting Information Figure S1). The regional MF values observed here were consistent with our previous 3T study, which reported MF of approximately 9-10% in the corpus callosum and corona radiata[23]. Rosette UTE and VFA-based relaxometry reported approximately 6-11% MF across major white-matter regions[37,38].

MTsat and MTR were derived from the same MPM acquisitions, but only MTsat remained consistent across field strengths. The MPM protocols were optimized separately for each scanner and differed in repetition time and excitation flip angle (Table 1). Because MTR depends directly on the MT-pulse duty cycle and the flip angle, these protocol differences, rather than the field strength itself, likely explain most of the systematic MTR offset between 1.5T and 3T. MTsat accounts for these dependencies through its explicit $T_1$ and $B_1$ correction[5,9], and the MT saturation of the semisolid pool at a fixed off-resonance frequency is expected to depend only weakly on $B_0$. The cross-field results thus support the use of correction-based MT metrics such as MTsat when data from different scanners and protocols are combined. Multi-UTE MF is an apparent ultrashort-$T_2^*$ signal fraction shaped by component-specific weighting and the sequence response. A common receive-sensitivity factor cancels in the fraction, whereas $B_1^+$ variation changes the effective flip angle and contribute to cross-field MF differences. Additionally, the multicomponent fit depends on the minimum achievable TE relative to the ultrashort-$T_2^*$ distribution, on readout-related blurring, on SNR, and on $T_1$ weighting that is not included in the current fitting procedure. All of these vary between field strengths and are discussed below.

Other multi-field-strength comparisons of myelin-sensitive metrics have reported similarly mixed results. MWF has been shown to correlate strongly between 1.5T and 3T ($R^2$>0.8) but with systematically higher absolute values at 3T, likely reflecting SNR- and T2-dependent fitting bias rather than true tissue differences. MPM studies have demonstrated that MT- and PD-based metrics show relatively small field-strength dependence, whereas $T_1$ and $T_2^*$ are more strongly field-dependent, and have cautioned that direct pooling of quantitative maps across field strengths remains complicated without dedicated relaxometry correction models. These prior studies[9,39], like the present one, generally involved small cohorts and restricted scanner limiting generalizability across scanner platforms and field-strength combinations.

Across all regions, MF and MTsat were good correlated at both field strengths (FreeSurfer atlas: $R^2$ = 0.79 at 3T and 0.73 at 1.5T; Supporting Information Figure S6). This correlation largely reflects the shared contrast between myelin-rich white matter and myelin-poor gray matter. Within white matter (JHU atlas), the correlation was lower ($R^2$ = 0.65 at 3T and 0.43 at 1.5T), partly because the inter-tract range of both metrics is narrower than the WM–GM contrast. At 3T, the MF test–retest $R^2$ across the same tracts was 0.97–0.99 (Table 2, Fig. 5), so MF measurement noise accounts for only a small part of the unshared variance. The remainder reflects either MTsat measurement variability, which could not be quantified because test–retest data were not acquired for MTsat, or genuine differences between the two metrics. At 1.5T, MF repeatability was lower ($R^2$ = 0.89–0.92), which may explain part, but not all, of the weaker correlation. The difference in the information provided by the MF and MTsat can be better understood in pathology, where demyelination, inflammation, and edema are expected to affect the two metrics differently.

Susceptibility effects and fiber orientation may contribute to discrepancies between multi-UTE-derived MF and the underlying myelin content. In iron-rich gray matter, iron shorten apparent $T_2^*$, while rapidly decaying signals may also arise from non-myelin cellular membranes, potentially biasing MF upward[40–42]. In white matter, the susceptibility anisotropy of myelin makes $T_2^*$ dependent on fiber orientation relative to $B_0$, suggesting an orientation dependence of MF that remains to be quantified[43,44]. Thus, MF should be considered myelin-sensitive rather than myelin-specific. Because the $R_2^*$ maps from the MPM protocol and the $B_0$ maps from the dual-echo multi-UTE data are available without additional acquisition time, iron-related bias could be assessed by including these measures as covariates in future analyses. The $B_0$ maps derived from the two multi-UTE echoes (Supporting Information Figure S3) showed better field homogeneity at 1.5T than at 3T. The $\Delta B_0$ range was approximately 150 Hz at 3T and about half of that at 1.5T in one healthy volunteer. The maximum $B_0$-induced phase accrual over ΔTE = 180 µs was smaller than 10°. Off-resonant protons associated with myelin lipids and proteins did not contribute significantly to the total phase because of their small signal fraction[16,19].

Several factors may account for the residual cross-field differences in MF. The myelin-associated non-aqueous proton signal is spectrally heterogeneous rather than arising from a single monoexponential pool. Ex vivo studies of isolated myelin and myelinated nerve identified multiple rapidly decaying components with a broad super-Lorentzian $T_2^*$ distribution extending from approximately 8 µs to 26 ms[45]. Subsequent ex vivo human-brain experiments resolved an ultrashort-$T_2^*$ component attributed primarily to the myelin bilayer ($T_{2,min}^* \approx 5.5$ µ$s$) and a distinct short non-aqueous component[19]. The MF estimated with our clinical multi-UTE implementation is therefore best interpreted as an effective, acquisition- and model-dependent fraction of the detectable ultrashort-$T_2^*$ signal rather than a complete measure of all myelin-bound protons[19,37]. On clinical systems, the minimum TE is limited by RF-pulse duration, transmit-receive switching, and gradient performance. In addition, $T_2^*$ decay during the radial readout broadens the effective point-spread function, causing signal loss and spatial blurring[46,47]. Because $T_2^*$ depends on $B_0$[48], nominally identical acquisition parameters may capture different portions of the underlying signal distribution at 1.5T and 3T. The lower SNR at 1.5T may increase noise sensitivity during multicomponent fit, which is consistent with the lower and more variable ICC values than 3T. Incomplete sampling of the distributed ultrashort-$T_2^*$ signal, together with $T_1$ and $T_2^*$ weighting, may explain higher MF obtained with TE distribution 2, and field-dependent PSF broadening may alter spatial signal mixing in the major central white matter tracts. Finally, the current signal model does not account for $T_1$ weighting, so variations in the water-pool signal and $T_1$ relaxation may be absorbed into the fitted component fractions. This may also contribute to the systematic between-day offset of -0.0031 observed at 3T, together with physiological variability such as changes in hydration, which was not measured. Incorporating $T_1$ information into the model may reduce this bias (Supporting Information Figure S7), although reliable estimation of component-specific $T_1$ values is difficult because the signals of different tissue pools are strongly coupled within each voxel at the SNR available for ultrashort-$T_2^*$ imaging. With a total acquisition time of 28 min, multi-UTE is considerably faster than earlier ultrashort-$T_2^*$ mapping protocols, such as the 50-minute VFA protocol by Deveshwar et al.[37] and the 90-minute HYFI protocol by Baadsvik et al.[19]. However, it remains too long for routine clinical use and is longer than the MPM protocol used here, which additionally provides $R_1$, $R_2^*$, and PD maps. Distributing the TEs across neighboring spokes averages motion effects over the echo series but cannot eliminate residual motion, which may introduce echo-dependent signal inconsistencies and bias the voxel-wise multicomponent fit. Increasing the number of spokes improves SNR and fit stability but prolongs the scan and increases the susceptibility to motion. We have recently shown that dual-domain self-supervised deep-learning reconstruction enables a 5-fold acceleration of multi-UTE while preserving MF estimation[49], which may help resolve this trade-off.

This study has several limitations: First, the cohort was small (n=5), repeatability was assessed in only two subjects, and TE distribution 2 at 1.5T was acquired only in these two subjects, which limits statistical power and generalizability, particularly to patients with demyelinating disease. Second, each field strength was represented by a single scanner from one vendor, so field strength is confounded with scanner hardware, and inter-vendor or multi-site variability, against which MTsat has previously been validated[8,9], was not assessed. Third, the MPM protocols were not matched between field strengths, $B_1$ inhomogeneity was corrected with the data-driven UNICORT approach rather than a measured $B_1$ map, and test–retest data were acquired only for multi-UTE, not for MTsat. Fourth, ROI-wise agreement metrics depend on the inter-regional dynamic range of each metric, and residual registration errors between the MF maps and the atlases may increase variability in small ROIs despite the fully automated pipeline. Finally, the current signal model does not account for $T_1$ weighting or the full spectral heterogeneity of the ultrashort-$T_2^*$ pool.

# 5. CONCLUSION

Multi-UTE enables quantitative mapping of the ultrashort-$T_2^*$ signal fraction on clinical 1.5T and 3T systems, together with high-resolution structural images and $B_0$ maps, in a single acquisition. The multi-UTE-derived myelin fraction showed high repeatability at 3T and cross-field-strength reproducibility comparable to that of MTsat, without requiring $B_1$ or $T_1$ mapping on clinical MRI systems in a small cohort of healthy volunteers. MF and MTsat were strongly correlated across brain regions, largely driven by the WM-GM contrast, whereas the positive intercept and steeper inter-tract slope of MF indicate that the two metrics are not interchangeable. Reducing the acquisition time, for example through accelerated acquisition and deep learning-based reconstruction, together with improved SNR at 1.5T, inclusion of $T_1$ weighting in the signal model, and validation in patients and multi-site settings will be important steps toward the use of multi-UTE in longitudinal and multicenter studies of myelin-related tissue changes.

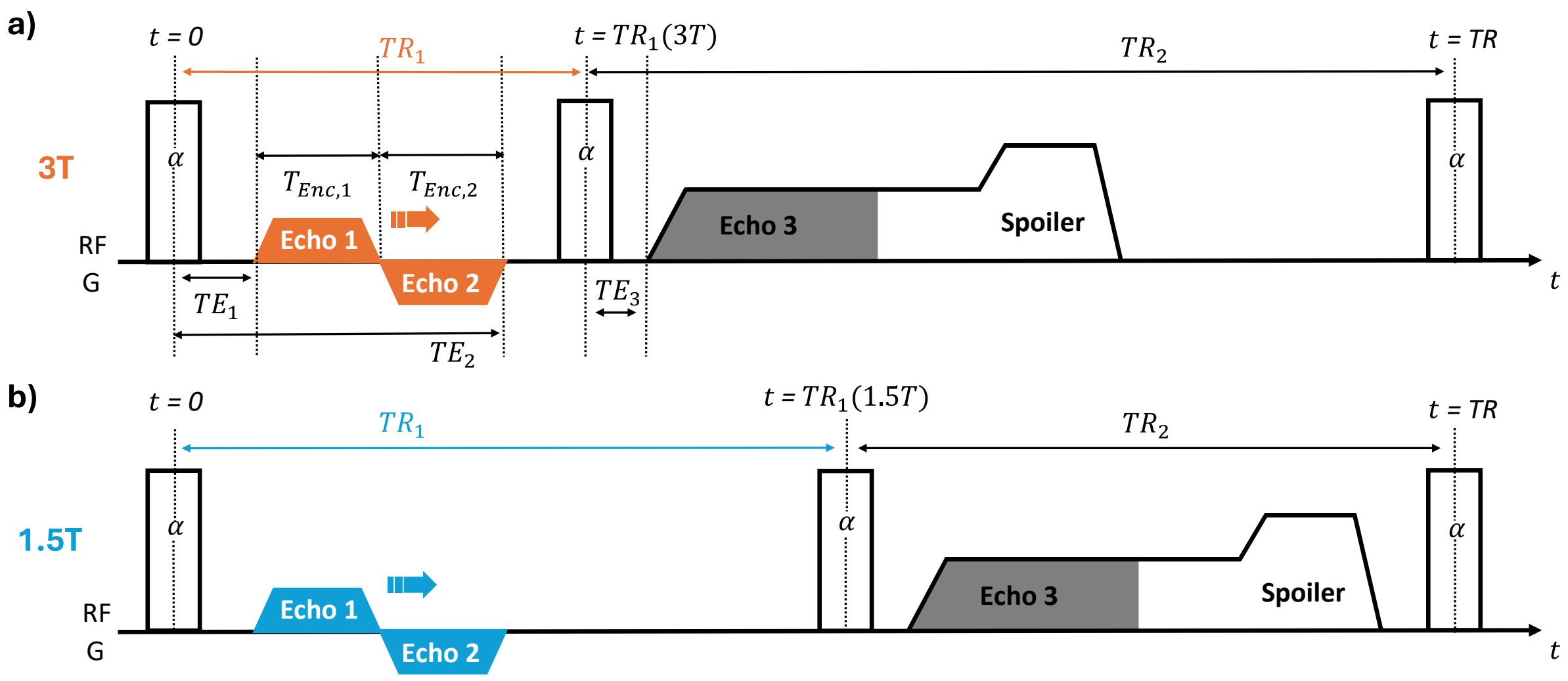

***Figure 1: Sequence diagram of the multi-contrast dual-resolution UTE (multi-UTE) at 3T and 1.5T. (a)*** *A 1-1 binomial water excitation pulse (FA=4°, $T_{RF}$=20µs, inter-pulse delay $TR_1$ = 1.19ms) was used at 3T. The first two echoes ($S_1$ and $S_2$) were acquired using a bipolar gradient with zero gradient moment. Consequently, the corresponding k-space trajectory started and ended at the center of k-space. Spoke-dependent echo times ($TE_1$ and $TE_2$) were used for low-resolution $S_1$ and $S_2$, whereas high-resolution $S_3$ was acquired at a fixed echo time $TE_3$ and reconstructed using all acquired spokes, resulting in number of spokes $N_{S3}$ = $N_{acq}$* $N_{S1}$.* ***(b)*** *At 1.5T, $TR_1$ was doubled relative to that at 3T ($TR_1$ = 2.38ms) to account for the field-dependent water–fat frequency separation, while also providing a broader TE-encoding range for $S_1$ and $S_2$.*

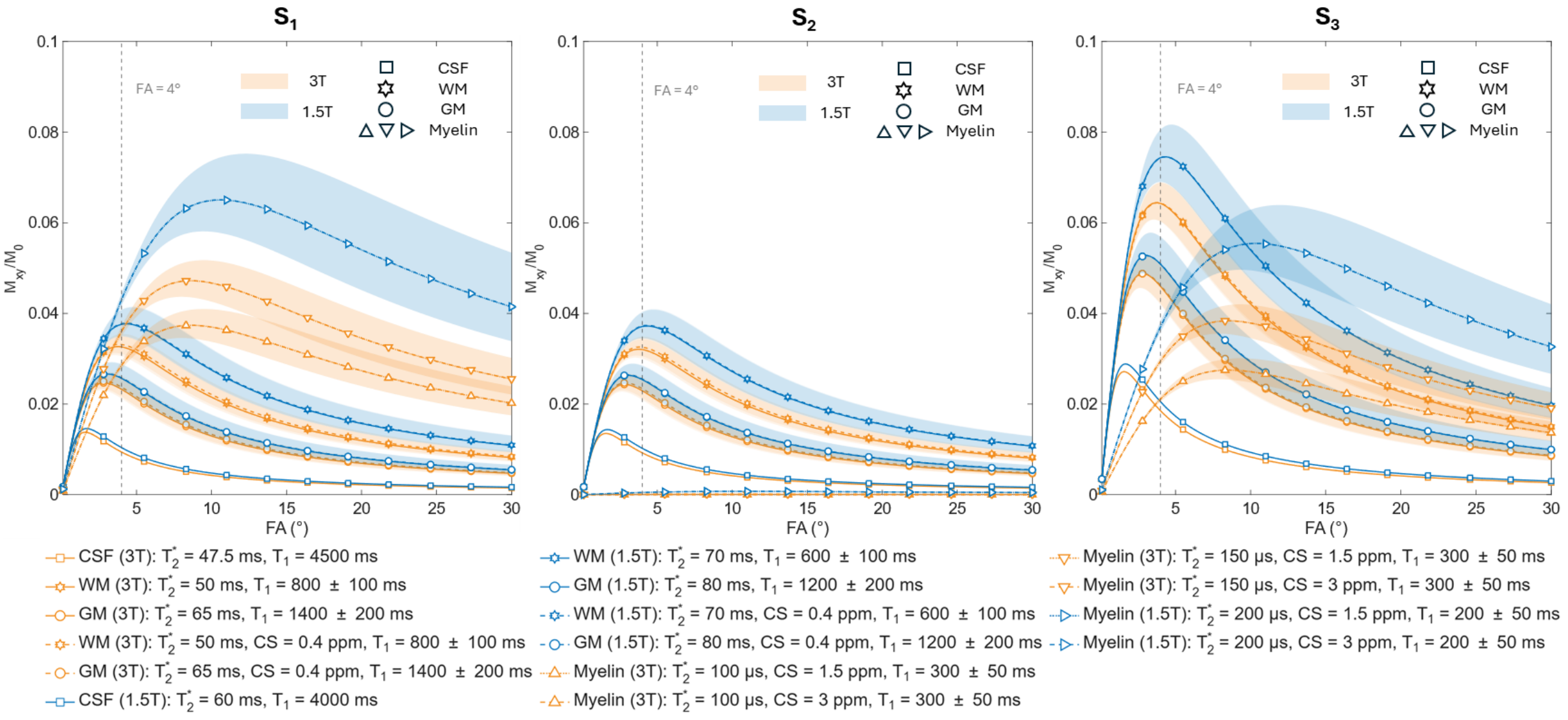

***Figure 2: Simulated signal curves for multi-UTE at 3T and 1.5T**. The normalized transverse magnetization ($M_{xy}/M_0$) was simulated as a function of flip angle for cerebrospinal fluid (CSF), white matter (WM), gray matter (GM) and myelin at 3T (orange) and 1.5T (blue). Field-strength-dependent $T_1$ values and the effects of different chemical shifts (CS) and $T_2^*$ values were incorporated. The three panels show the simulated signal responses for $S_1$ ($TE_1$=70μs), $S_2$ ($TE_2$=970μs) and $S_3$ ($TE_3$=100μs), respectively. The vertical dashed line marks the flip angle used in the experiments (FA=4°). The $T_1$, $T_2^*$ and chemical-shift values used for each tissue component and field strength are listed below the plots.*

| Parameter | multi-UTE | | T1_MPRAGE | | MPM | | | | | |
|---|---|---|---|---|---|---|---|---|---|---|
| | | | | | T1w_FL3D | | PDw_FL3D | | MTw_FL3D | |
| **Field strength** | **3T** [a] | **1.5T** | **3T** | **1.5T** | **3T** | **1.5T** | **3T** | **1.5T** | **3T** | **1.5T** |
| **Acquisition time (min:s)** | 28:07 | 28:07 | 4:05 | 4:09 | 4:04 | 5:02 | 6:17 | 8:07 | 6:17 | 8:07 |
| **Voxel size ($mm^3$)** | 1.7x1.7x1.7;<br>0.5×0.5×0.5 | 1.7x1.7x1.7;<br>0.5×0.5×0.5 | 1.0×1.0×1.0 | 1.0×1.0×1.0 | 1.0×1.0×1.0 | 1.0×1.0×1.0 | 1.0×1.0×1.0 | 1.0×1.0×1.0 | 1.0×1.0×1.0 | 1.0×1.0×1.0 |
| **FOV** | 224x224x224 | 224x224x224 | 224x224x224 | 224x224x224 | 224x224x224 | 250x250x224 | 224x224x224 | 250x250x224 | 224x224x224 | 250x250x224 |
| **Phase-encoding direction** | None | None | A >> P | A >> P | A >> P | A >> P | A >> P | A >> P | A >> P | A >> P |
| **TR (ms)** | 7.0 | 7.0 | 2050.0 | 2200.0 | 22.0 | 31.0 | 34.0 | 50.0 | 34.0 | 50.0 |
| **TE (ms)** | $TE_1$ = { 0.07, 0.09, 0.11, 0.14, 0.17, 0.25};<br>$TE_2 = TE_1 + 0.9$;<br>$TE_3 = 0.10$ | $TE_{1,dist1}$ = {0.07, 0.09, 0.11, 0.14, 0.17, 0.25} [b];<br>$TE_{1,dist2}$ = { 0.07, 0.11, 0.15, 0.19, 0.31, 0.55} [b];<br>$TE_2 = TE_1 + 0.9$;<br>$TE_3 = 0.10$ | 2.39 | 2.52 | {1.92, 4.25, 6.59, 8.93, 11.27, 13.61, 15.95, 18.29} | {2.12, 6.63, 13.00, 19.37, 25.74} | {2.30, 4.73, 7.19, 9.65, 12.11, 14.57, 17.03, 19.49} | {2.12, 6.63, 13.00, 19.37, 25.74, 32.11, 38.48, 44.85} | {2.30, 4.73, 7.19, 9.65, 12.11, 14.57} | {2.12, 6.63, 13.00, 19.37, 25.74} |
| **Flip angle (°)** | 4/4 [c] | 4/4 [c] | 8 | 15 | 21 | 20 | 4 | 5 | 4 | 5 |
| **Magn. preparation / TI** | None | None | Non-sel. IR;<br>TI = 900 ms | Non-sel. IR;<br>TI = 973 ms | None | None | None | None | None | None |
| **Magnetization transfer contrast** | Off | Off | Off | Off | Off | Off | Off | Off | On | On |
| **Bandwidth (Hz/Px)** | 2222/1449 [d] | 2222/1449 [d] | 260 | 160 | 490/480 | 480/160 | 490/480 | 480/160 | 490/480 | 480/160 |
| **Special parameters** | $T_{RF}$ = 20 µs;<br>$TR_1$ = 1190 µs;<br>$G_1 = G_2$ = 20 mT/m;<br>$G_3$ = 36 mT/m;<br>SR = 180 mT/m/ms;<br>$N_{spokes,S1} = N_{spokes,S2}$ = 40000;<br>$N_{spokes,S3}$ = 240000 | $T_{RF}$ = 20 µs;<br>$TR_1$ = 2380 µs;<br>$G_1 = G_2$ = 20 mT/m;<br>$G_3$ = 36 mT/m;<br>SR = 180 mT/m/ms;<br>$N_{spokes,S1} = N_{spokes,S2}$ = 40000;<br>$N_{spokes,S3}$ = 240000 | | | | | | | | |

[a] For test–retest measurements, multi-UTE were repeated five times in subject 1 and three times in subject 2.
[b] For test–retest measurements, multi-UTE using two TE distributions were each repeated three times in subject 1 and twice in subject 2. For the remaining subjects, the TE distribution 1 were used for cross-field-strength comparisons.
[c] Multi-UTE acquisitions were performed using two hard pulses, each with a flip angle of 4°.
[d] The bandwidth correspond to the low-resolution multi-UTE image ($S_1$ and $S_2$) and the high-resolution multi-UTE image ($S_3$), respectively.

***Table 1: MRI acquisition parameters at 3T and 1.5T.*** *Protocol settings are summarized for multi-UTE, T1-weighted magnetization-prepared rapid gradient-echo (T1w-MPRAGE) and multiparameter mapping (MPM), including T1-weighted, proton-density-weighted and magnetization-transfer-weighted three-dimensional fast low-angle shot (T1w-, PDw- and MTw-FL3D) acquisitions.*

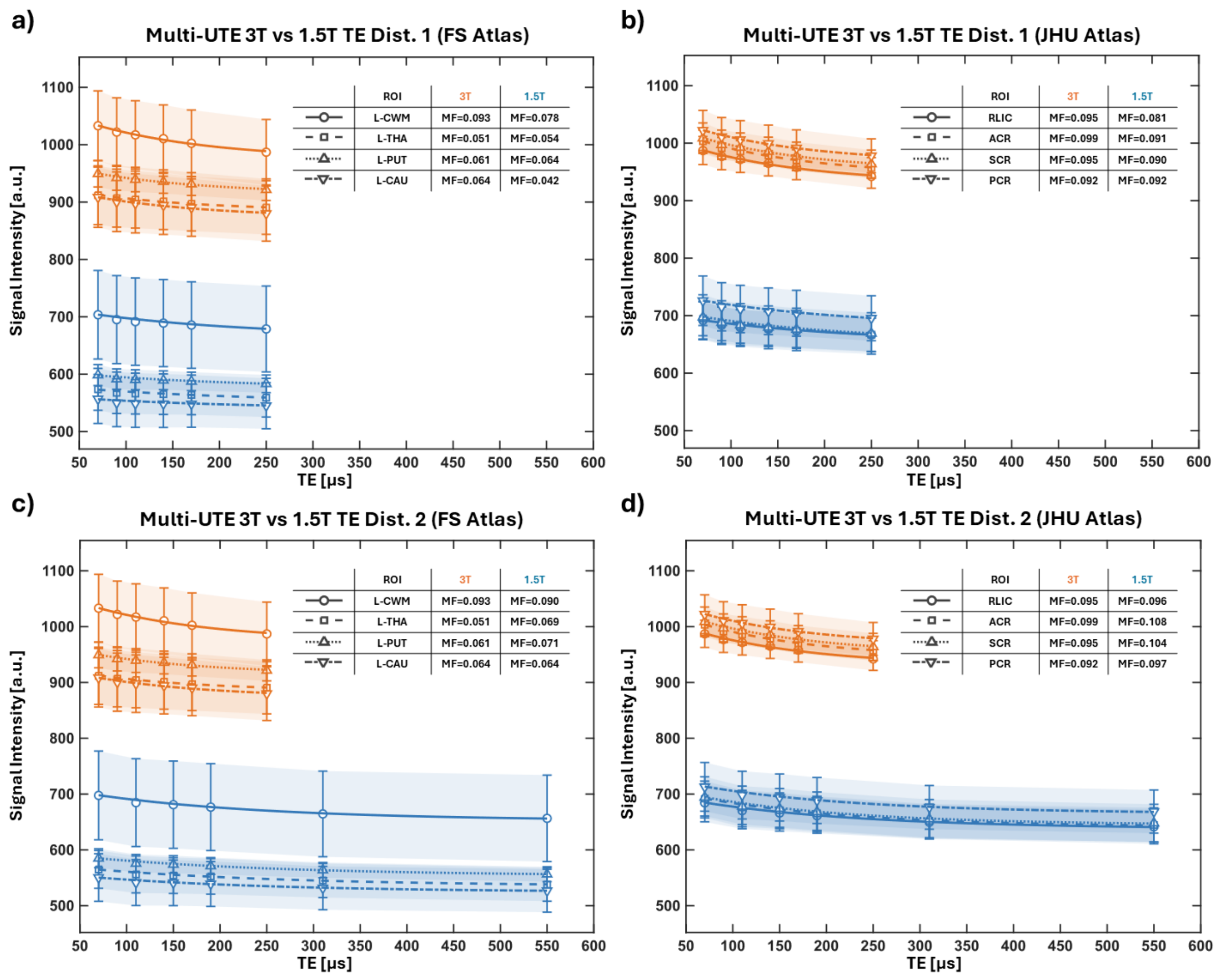

***Figure 3: In vivo signal behavior of multi-UTE in selected ROIs at 3T (orange) and 1.5T (blue).*** *Signal intensity (a.u.) is plotted against echo time (TE, μs).* ***(a,b)*** *multi-UTE 3T data compared with 1.5T data acquired using TE distribution 1.* ***(c,d)*** *Corresponding comparison using TE distribution 2 at 1.5T. Panels a and c show FreeSurfer-defined ROIs: left cerebral white matter (L-CWM), thalamus (L-THA), putamen (L-PUT), and caudate (L-CAU). Panels b and d show selected JHU DTI atlas tracts: the retrolenticular part of the internal capsule (RLIC) and the anterior, superior, and posterior corona radiata (ACR, SCR, and PCR). Error bars show ±1 SD across voxels. Insets report estimated MF values.*

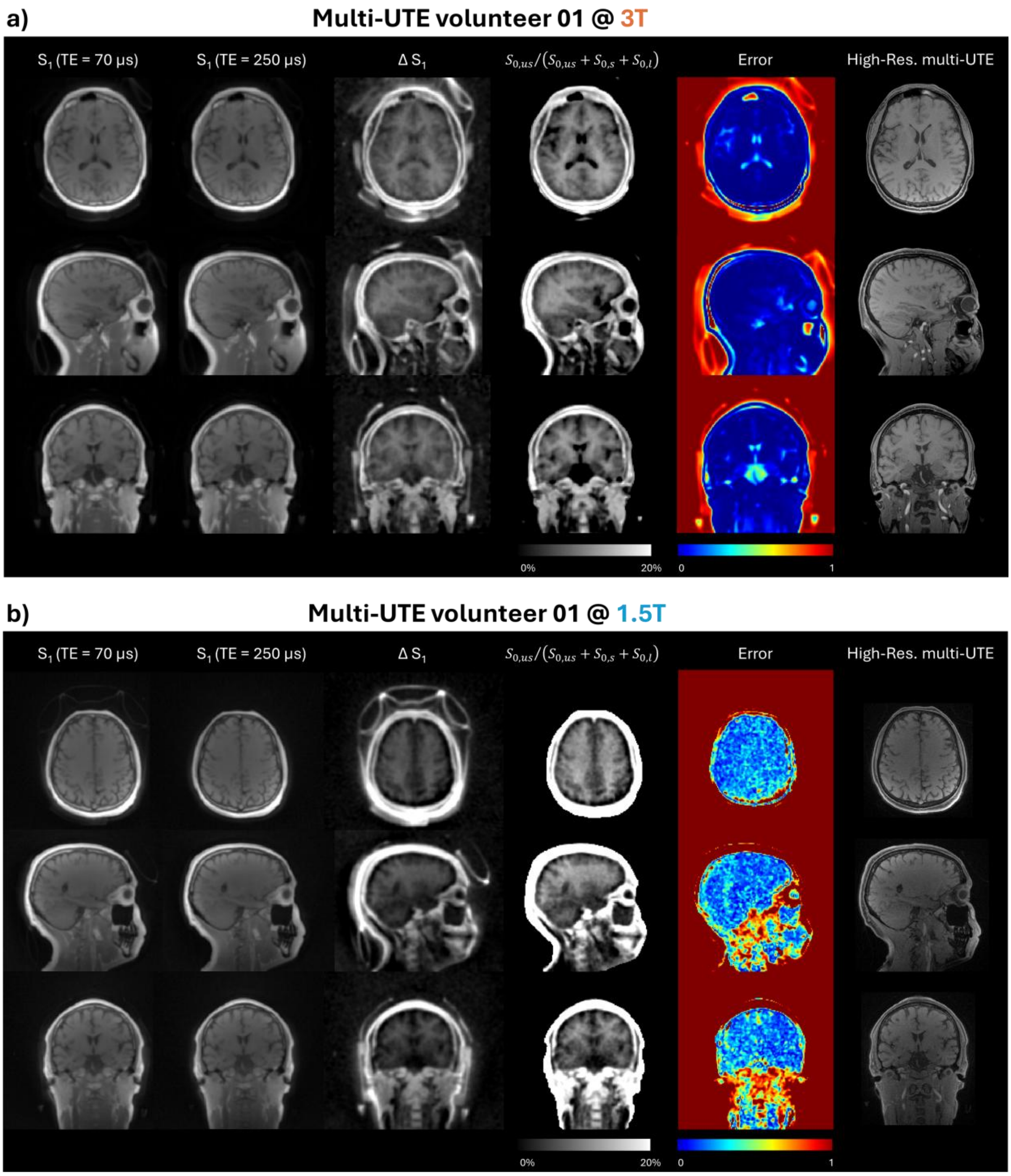

***Figure 4: In vivo multi-UTE results from a healthy volunteer.*** ***(a)*** *Results at 3T.* ***(b)*** *Results at 1.5T. From left to right:* $S_1(TE = 70\ \mu s)$, $S_1(TE = 250\ \mu s)$, $\Delta S_1$, *ultrashort-*$T_2^*$ *signal fraction map (* $S_{0,us}/\left(S_{0,us} + S_{0,s} + S_{0,l}\right)$ *), estimation-error map and high-resolution multi-UTE* $S_3(TE_3 = 100\ \mu s)$.

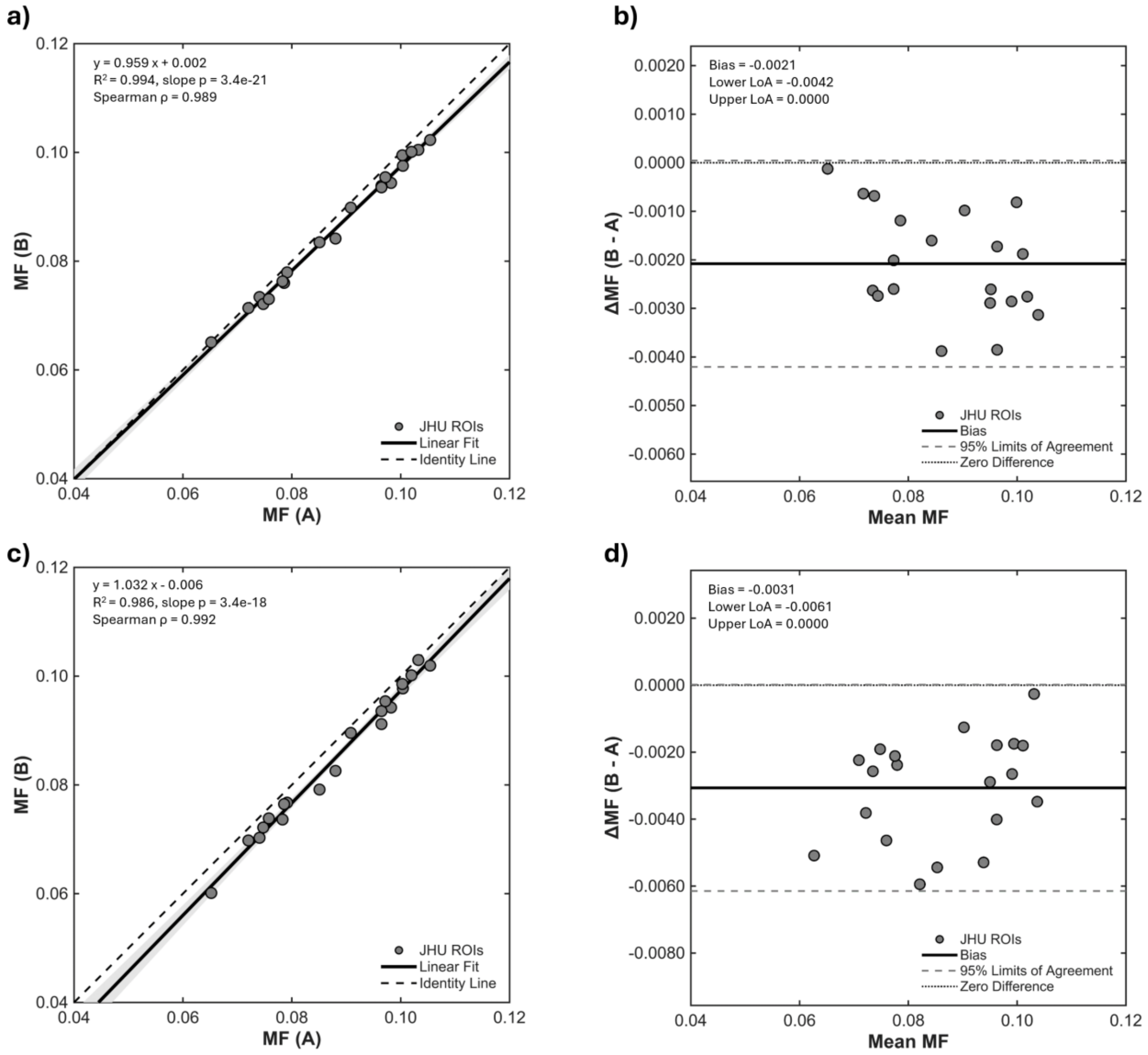

***Figure 5: Same-day and between-day repeatability of MF using multi-UTE at 3T.*** *The same volunteer (subject 1) was scanned using the multi-UTE protocol with TR = 7 ms, twice within the same day (A: scan 2, B: scan 3) and once on a different day (A: scan 2, B: scan 4).* ***(a, b)*** *Linear-regression and Bland–Altman analyses of the same-day scans are shown.* ***(c, d)*** *Corresponding analyses of scans acquired on different days. Additional results comparing regional correlations based on two different Atlases are presented in the Supporting Information Figure S4. All repeatability results of subject 1 are presented in the Supporting Information Figure S5.*

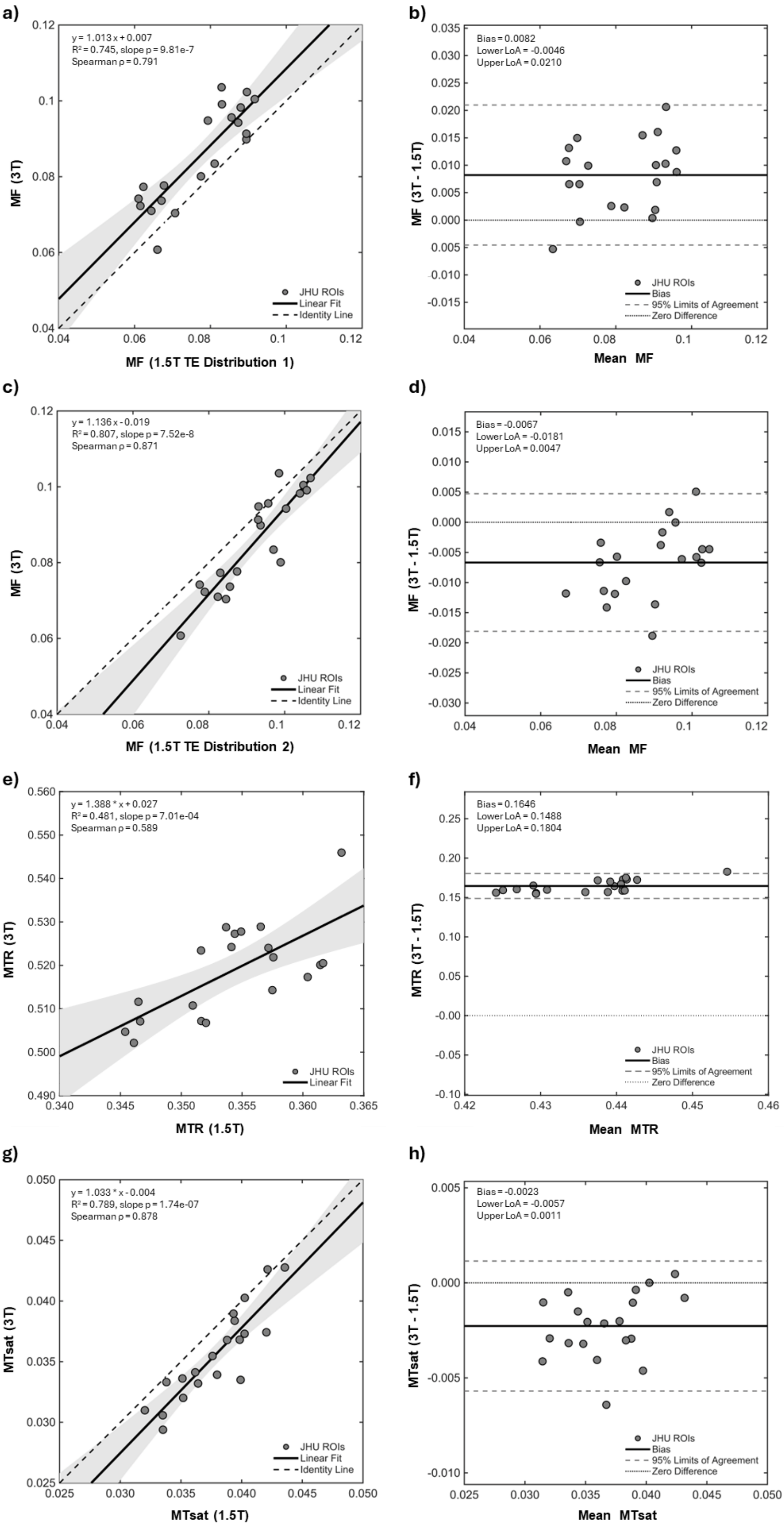
a)
y = 1.013 x + 0.007
R² = 0.745, slope p = 9.81e-7
Spearman ρ = 0.791
JHU ROIs
Linear Fit
Identity Line
MF (3T)
MF (1.5T TE Distribution 1)
b)
Bias = 0.0082
Lower LoA = -0.0046
Upper LoA = 0.0210
JHU ROIs
Bias
95% Limits of Agreement
Zero Difference
MF (3T - 1.5T)
Mean MF
c)
y = 1.136 x - 0.019
R² = 0.807, slope p = 7.52e-8
Spearman ρ = 0.871
JHU ROIs
Linear Fit
Identity Line
MF (3T)
MF (1.5T TE Distribution 2)
d)
Bias = -0.0067
Lower LoA = -0.0181
Upper LoA = 0.0047
JHU ROIs
Bias
95% Limits of Agreement
Zero Difference
MF (3T - 1.5T)
Mean MF
e)
y = 1.388 * x + 0.027
R² = 0.481, slope p = 7.01e-04
Spearman ρ = 0.589
JHU ROIs
Linear Fit
MTR (3T)
MTR (1.5T)
f)
Bias = 0.1646
Lower LoA = 0.1488
Upper LoA = 0.1804
JHU ROIs
Bias
95% Limits of Agreement
Zero Difference
MTR (3T - 1.5T)
Mean MTR
g)
y = 1.033 * x - 0.004
R² = 0.789, slope p = 1.74e-07
Spearman ρ = 0.878
JHU ROIs
Linear Fit
Identity Line
MTsat (3T)
MTsat (1.5T)
h)
Bias = -0.0023
Lower LoA = -0.0057
Upper LoA = 0.0011
JHU ROIs
Bias
95% Limits of Agreement
Zero Difference
MTsat (3T - 1.5T)
Mean MTsat

***Figure 6: Cross-field-strength agreement of MF using multi-UTE, MTR and MTsat. (a, b)*** *Linear regression and Bland–Altman analyses, respectively, comparing MF obtained at 3T using multi-UTE (TR = 7 ms) with MF obtained at 1.5T using multi-UTE TE distribution 1.* ***(c, d)*** *Corresponding analyses using the 1.5T TE distribution 2 acquisition.* ***(e, f)*** *Linear regression and Bland–Altman analyses, respectively, comparing MTR measured at 3T and 1.5T.* ***(g, h)*** *Corresponding analyses of MTsat. Each circle represents a region of interest (ROI) from the JHU DTI atlas. In the regression plots, the solid and dashed lines denote the linear fit and line of identity, respectively. In the Bland–Altman plots, the solid, dashed and dotted lines denote the mean bias, 95% limits of agreement and zero difference, respectively.*

| Metric (mean) | Test-retest Repeatability of multi-UTE MF (sub1 / sub2) | | | Cross-field-strength Reproducibility (5 subjects) | | |
|---|---|---|---|---|---|---|
| | 3T | 1.5T TE distribution 1 | 1.5T TE distribution 2 | multi-UTE MF | MTR | MTsat |
| **Slope** | 1.0/0.93 | 0.95/0.97 | 1.1/1.0 | 0.98 | 0.80 | 1.0 |
| Intercept | 0.00083/0.0060 | 0.0033/0.0012 | -0.0043/0.0077 | 0.0070 | 0.14 | -0.0032 |
| **$R^2$** | 0.98/0.97 | 0.89/0.92 | 0.95/0.91 | 0.68 | 0.23 | 0.77 |
| Mean bias | 0.0012/ 0.0011 | -0.0011/-0.0010 | 0.0019/0.0085 | 0.0046 | 0.072 | -0.0034 |
| RMSE | 0.0040/0.0031 | 0.0060/0.0047 | 0.0029/0.0088 | 0.0090 | 0.089 | 0.0040 |
| **RMSE / ROI SD** | 0.32/0.27 | 0.57/0.45 | 0.30/1.1 | 0.86 | 14 | 1.1 |
| Median absolute symmetric diff. (%) | 4.3/2.8 | 6.2/4.1 | 2.9/10 | 8.8 | 21 | 8.9 |
| **CCC** | 0.94/0.96 | 0.85/ 0.89 | 0.95/0.63 | 0.70 | 0.06 | 0.60 |
| **ICC** | 0.94/0.96 | 0.85/0.90 | 0.96/0.64 | 0.71 | 0.06 | 0.60 |

***Table 2. Test–retest repeatability of multi-UTE MF and cross-field-strength reproducibility of multi-UTE MF, MTR and MTsat.*** *Test–retest repeatability was assessed in two subjects; values correspond to sub1 and sub2, respectively. For each subject and acquisition protocol, test–retest metrics were computed separately for all unique pairs of repeated scans and then averaged across pairs. Figure 5 shows two individual scan pairs for subject 1 at 3T. Cross-field-strength reproducibility was assessed in five subjects. Individual 3T test–retest results for subject 1 are provided in the Supporting Information Figure S5.*

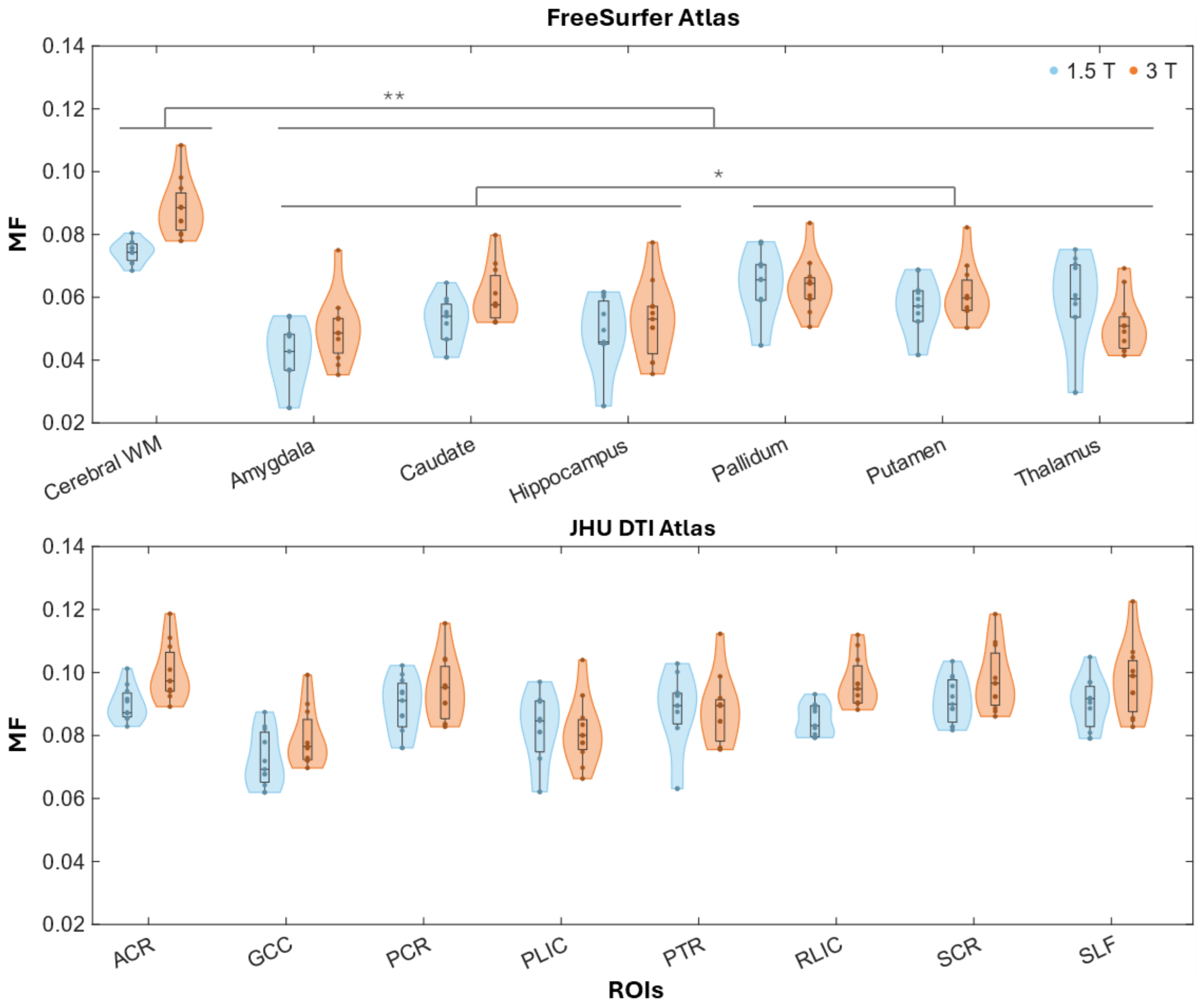

***Figure 7: Regional myelin fraction (MF) across healthy volunteers and acquisition protocols at 1.5T and 3T.*** *Violin plots show MF distributions for selected ROIs defined by the FreeSurfer atlas (top) and JHU DTI atlas (bottom), incorporating measurements from different volunteers and protocol settings. Blue and orange indicate 1.5T and 3T, respectively. Points represent individual measurements, and embedded box plots indicate the median and interquartile range. Statistical significance is denoted by * (P < 0.05) and ** (P < 0.01).*

# Multi-UTE Myelin Fraction Mapping: Repeatability and Cross-Field-Strength Reproducibility at 1.5T and 3T

Nan Yin[1], Marco Reisert[1], Theo Demerath[2], Shuai Liu[1], Alexander Rau[2], Uzay Emir[3,4], Horst Urbach[2], Michael Bock[1], Ali Caglar Özen[1]

[1]Division of Medical Physics, Department of Diagnostic and Interventional Radiology, University Medical Center Freiburg, Faculty of Medicine, University of Freiburg, Freiburg, Germany

[2]Department of Neuroradiology, Faculty of Medicine, University of Freiburg, Freiburg, Germany

[3]Department of Radiology, University of North Carolina, Chapel Hill, NC, USA

[4]The Lampe Joint Department of Biomedical Engineering, University of North Carolina, Chapel Hill, NC, USA

# SUPPORTING INFORMATION

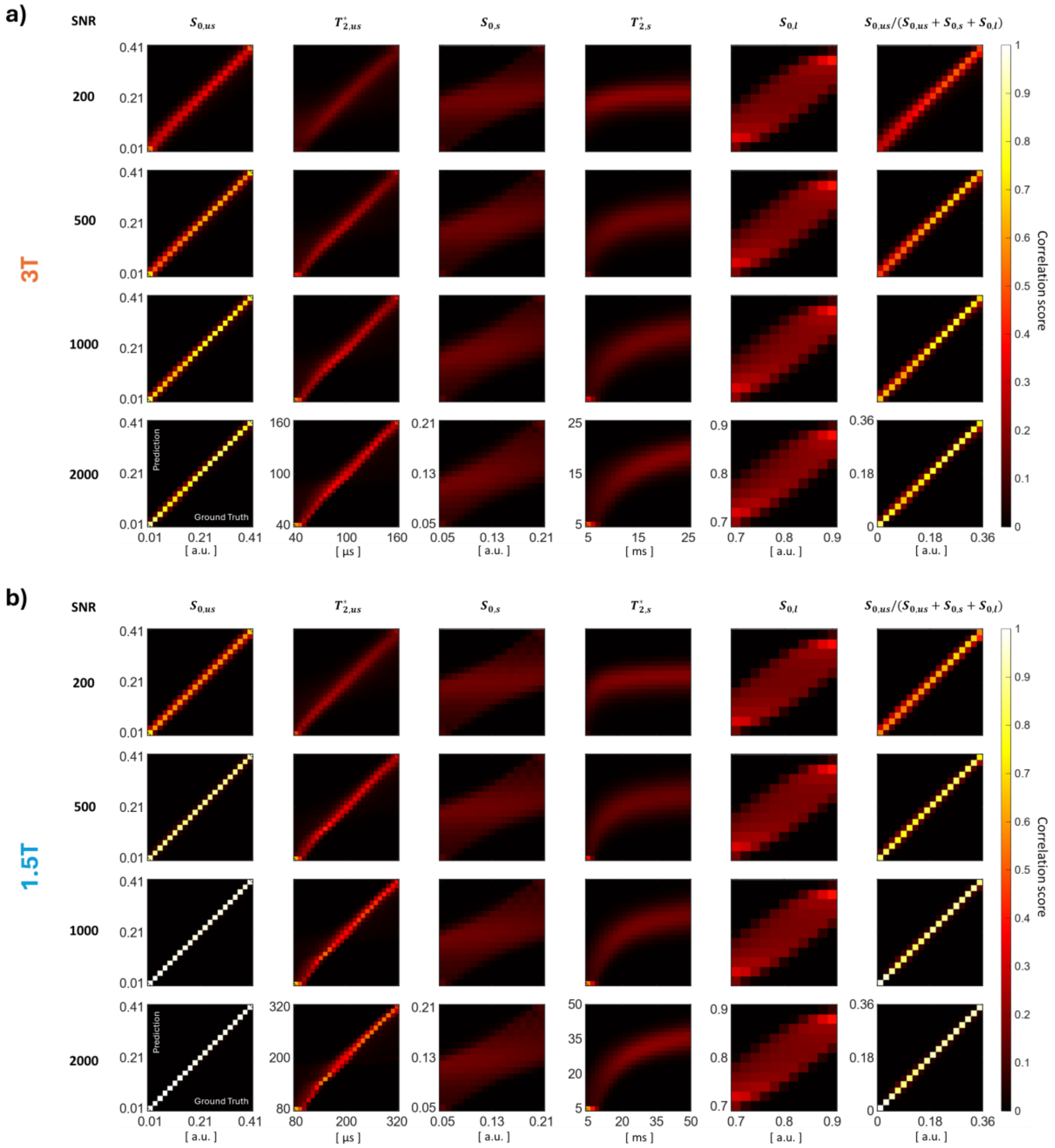

***Supporting Information Figure S1:*** *The correlation statistics of the Bayesian estimator for each used parameter of the signal model for SNR values between 200 and 2000 at 3T and 1.5T.*

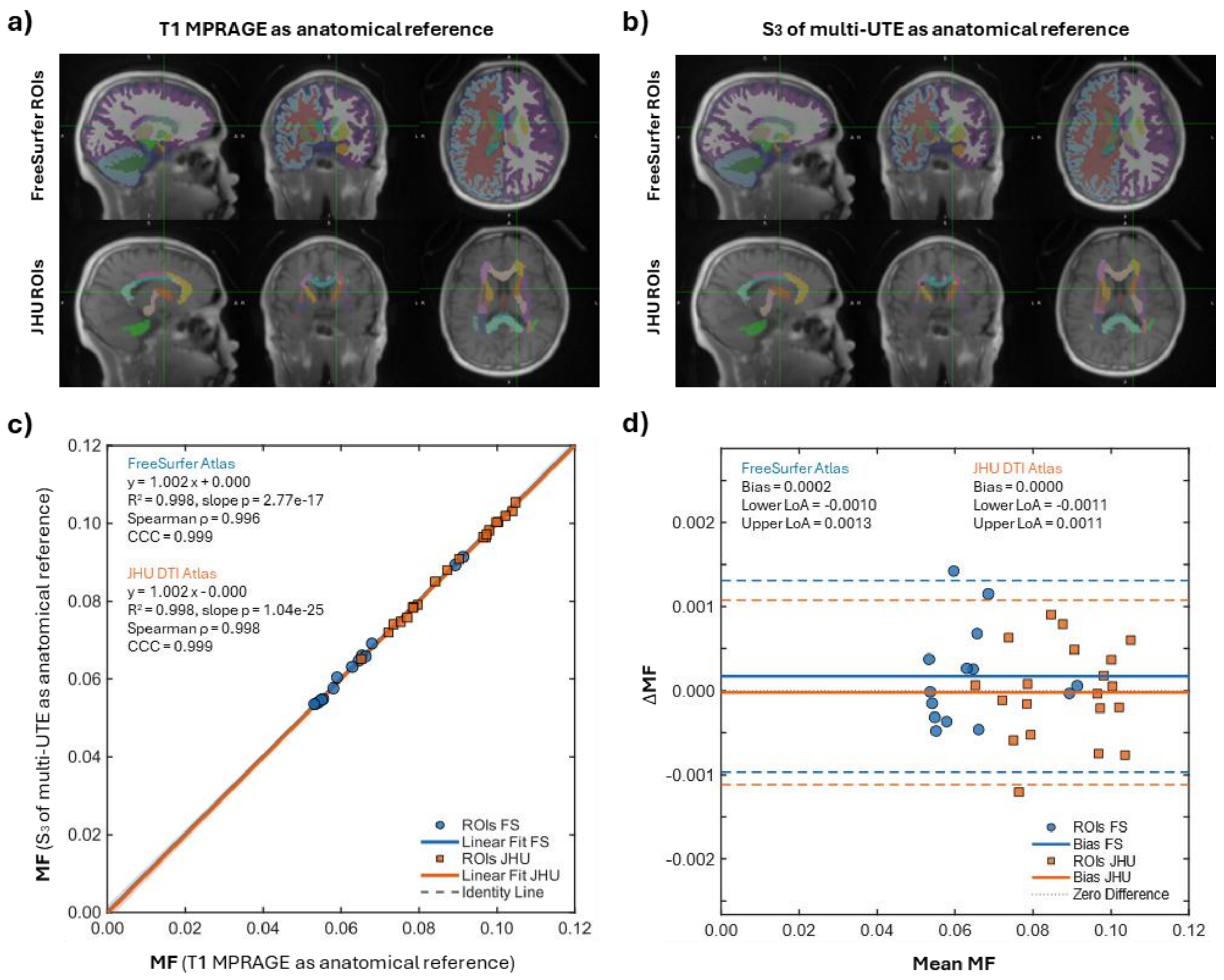

***Supporting Information Figure S2: Quality assessment of ROI segmentations generated using two anatomical references and agreement of the ROIs analysis results of the same scan. (a)*** *Representative sagittal, coronal, and axial views showing ROIs segmented using T1-weighted MPRAGE as the anatomical reference.* ***(b)*** *Corresponding ROIs segmented using $S_3$ of multi-UTE as the anatomical reference. In both panels, the segmentations are overlaid on the lower-resolution S1 image from the multi-UTE acquisition. The upper and lower rows show FreeSurfer and JHU DTI atlas ROIs, respectively. Colored regions denote individual ROIs, which were visually inspected for anatomical consistency and segmentation accuracy.* ***(c)*** *Linear regression comparing paired ROI-wise measurements obtained using the T1-MPRAGE as reference and $S_3$ of multi-UTE as reference segmentations.* ***(d)*** *Corresponding Bland–Altman analysis showing the mean bias, limits of agreement, and zero-difference line. Blue circles and orange squares represent FreeSurfer and JHU DTI atlas ROIs, respectively.*

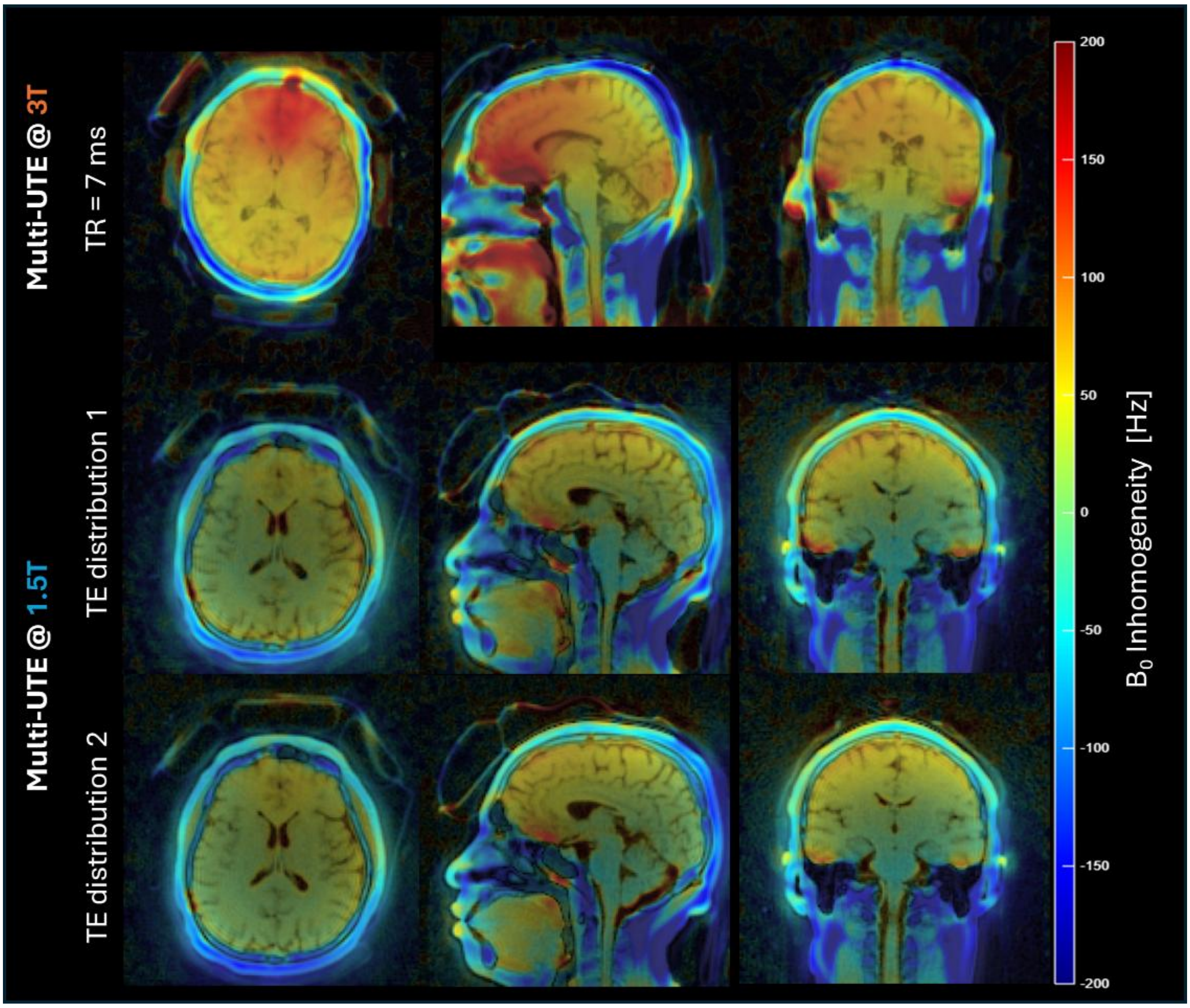

***Supporting Information Figure S3: $B_0$ maps obtained using multi-UTE of a healthy volunteer at 3T and 1.5T.*** *$B_0$ maps acquired on the 3T Prisma used TR = 7 ms; $B_0$ maps acquired on the 1.5T Sola used two TE-sampling schemes (TE distribution 1 and 2).*

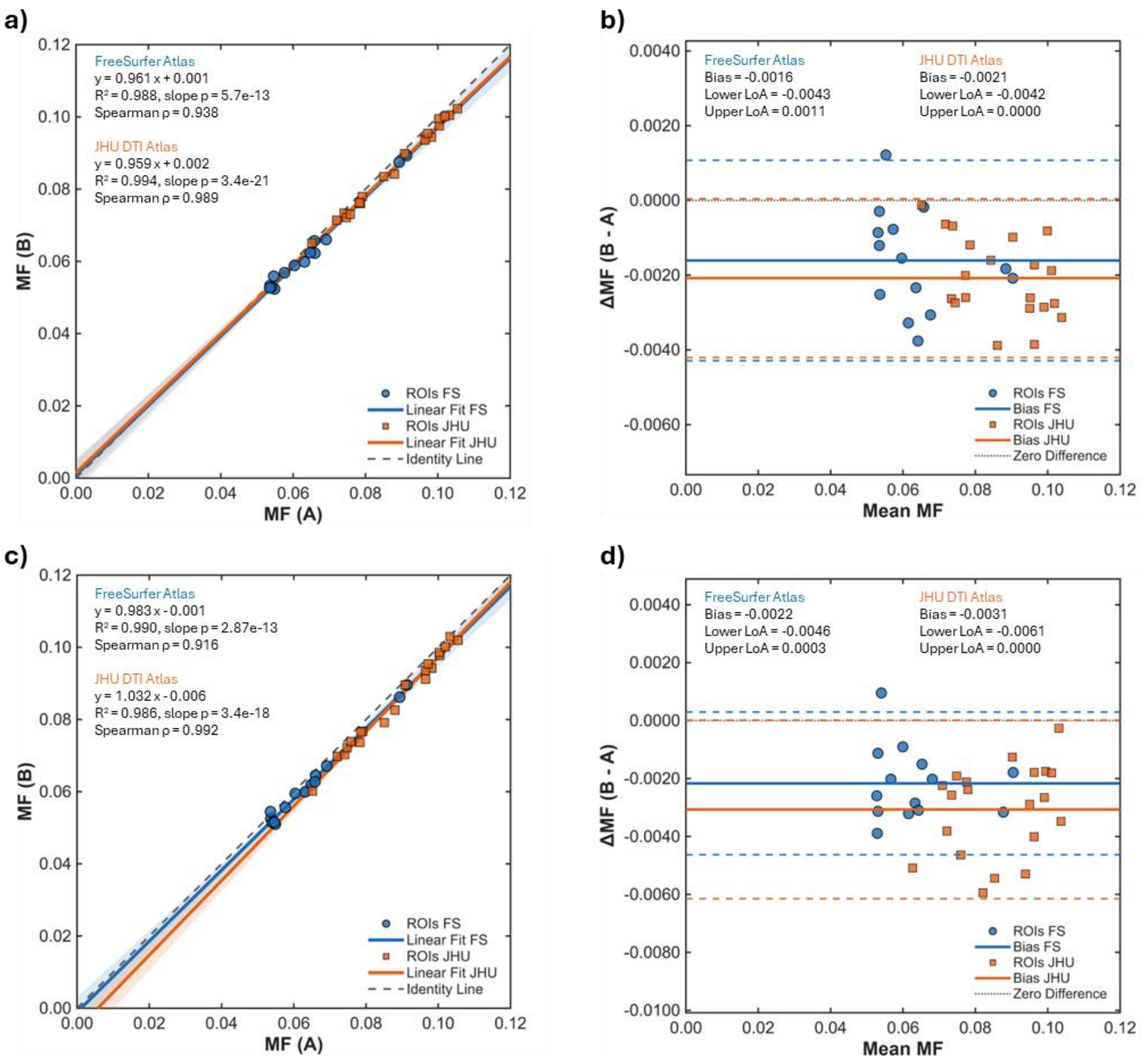

***Supporting Information Figure S4: Same-day and between-day repeatability of MF using multi-UTE at 3T. (a, b)*** *The same healthy volunteer (subject 1) was scanned using the same multi-UTE protocol with TR = 7 ms, both within the same day. Linear-regression and Bland–Altman analyses of the same-day scans (A: scan 2, B: scan 3), respectively. Blue circles and orange squares represent ROIs from the FreeSurfer and JHU DTI atlases, respectively.* ***(c, d)*** *Corresponding analyses of scans acquired on different days (A: scan 2, B: scan 4).*

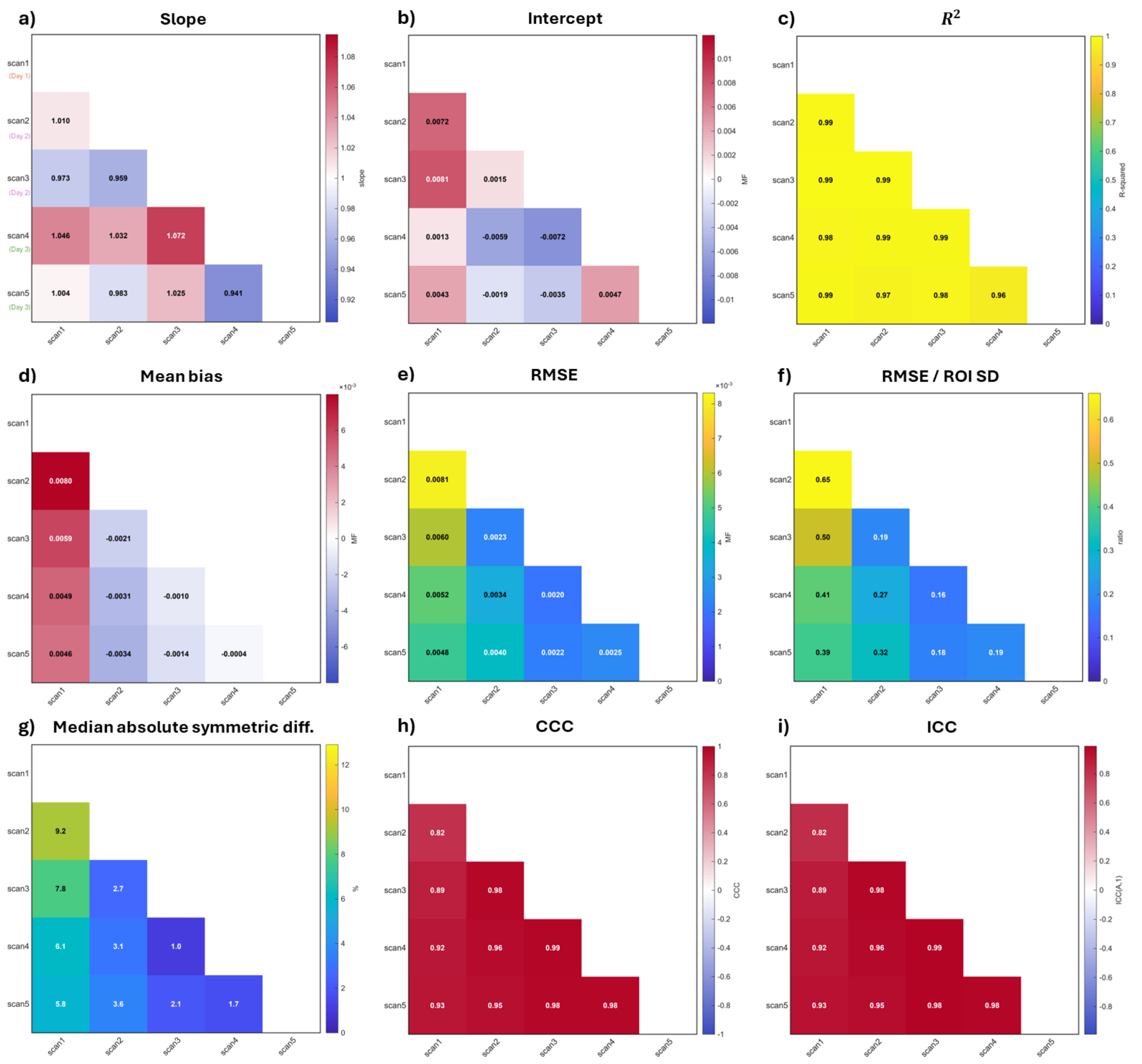

***Supporting Information Figure S5:*** *Pairwise agreement of regional MF using JHU atlas across five scans acquired over three days of one volunteer at 3T (multiUTE-TR7). Scan 1 was acquired on day 1, scans 2-3 on day 2, and scans 4-5 on day 3. From left to right and top to bottom, heatmaps show slope, intercept, $R^2$, mean bias, RMSE, RMSE/ROI SD, median absolute symmetric difference, CCC, and ICC.*

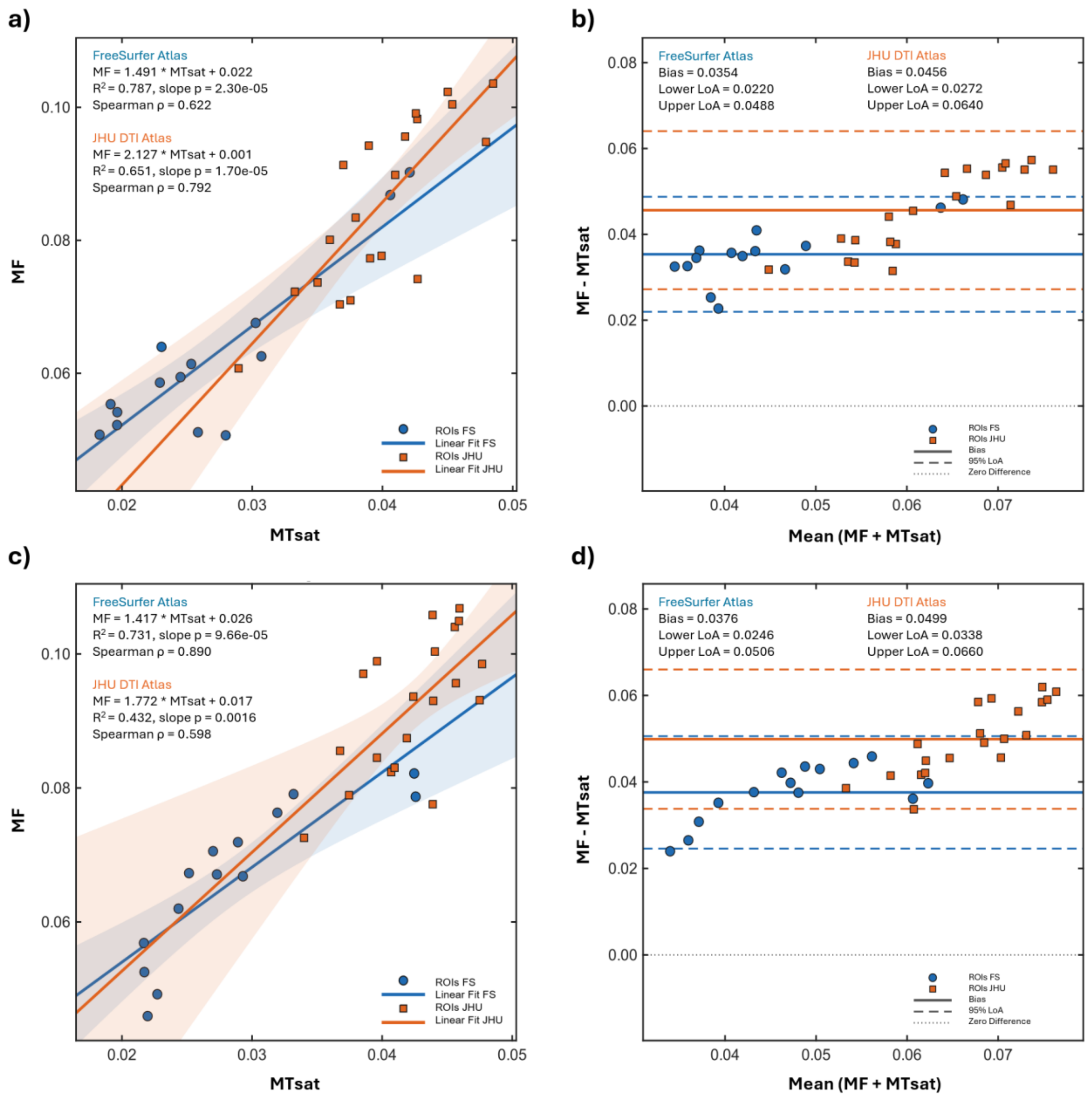

***Supporting Information Figure S6: Correlation between MTsat and multi-UTE derived MF at 1.5T and 3T. (a, b)*** *ROI-wise linear regression and Bland–Altman plot of MF vs MTsat at 3T.* ***(c, d)*** *ROI-wise linear regression and Bland–Altman plot of MF vs MTsat at 1.5T.*

## Simulated effects of field strength, $B_1^+$, and $T_1$ on multi-UTE MF estimates

Under the selected field-specific parameters, MF at FA = 4° was approximately 0.164 at 3T and 0.153 at 1.5T (input fraction of $\mathrm{M_0}$= 0.10). A ±20% variation in $\mathrm{B_1^+}$ produced approximately ±12% relative MF changes at both fields, decreasing to approximately ±6% under equal $\mathrm{T_1}$ conditions. At FA = 4°, unequal component $\mathrm{T_1}$ values increased MF by 24.8% at 3T and 21.4% at 1.5T relative to their matched equal $\mathrm{T_1}$ controls. Even with equal $\mathrm{T_1}$, MF remained above the input fraction of $\mathrm{M_0}$ =0.10 because component dependent binomial steady-state weighting persists after ideal readout-decay correction. These simulations identify both $\mathrm{B_1^+}$ variation and relaxation-dependent sequence weighting as potential contributors to differences in MF estimates across field strengths. Both $\mathrm{T_1}$ weighting and $\mathrm{B_1^+}$ variations influenced absolute MF values. However, the 3T/1.5T MF ratio was relatively insensitive to common $\mathrm{B_1^+}$ scaling, varying by less than 1% over the simulated ±20% range. These results suggest that matched $\mathrm{B_1^+}$ variations have a limited effect on relative cross-field comparisons, whereas the magnitude of the interfield MF difference remains dependent on the assumed $\mathrm{T_1}$ values.

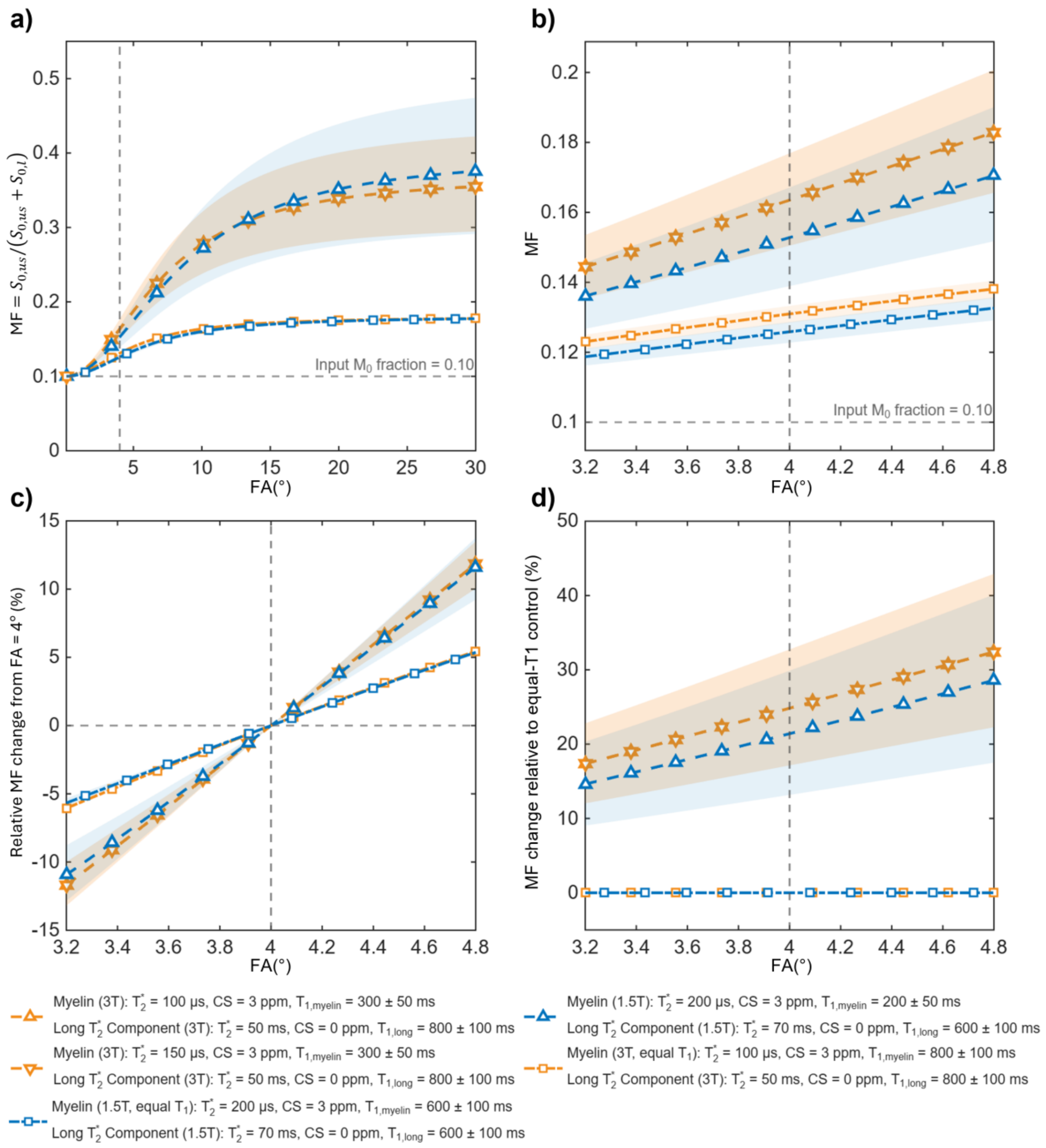

***Supporting Information Figure S7: Simulated effects of field strength, $B_1^+$, and $T_1$ on multi-UTE MF estimates.*** *Two components representing myelin and the long $T_2^*$ signal were simulated with $M_0$ fractions of 0.1 and 0.9. MF was calculated from amplitudes, assuming ideal fitting correction for $T_2^*$ decay.* ***(a)*** *multi-UTE MF equation of FA = 0.1–30°.* ***(b)*** *$\pm$ 20% $B_1^+$ inhomogeneity around a nominal FA of 4°.* ***(c)*** *Relative MF change from each condition's value at FA = 4°.* ***(d)*** *Percentage MF difference from a matched equal $T_1$ control at the same FA.*